\documentclass[a4paper]{article}
\usepackage[latin1]{inputenc} %
\usepackage{RR}
\usepackage{graphicx}    
\usepackage{listings}   
\usepackage{paralist}        
\usepackage{hyperref}
\hypersetup{ 
colorlinks,%
citecolor=black,%
filecolor=black,%
linkcolor=black,%
urlcolor=black
}  
\usepackage{macroRap} 

\RRdate{Decembre 2009}

\RRauthor{
Pierre Deransart\thanks{INRIA Paris-Rocquencourt, Pierre.Deransart@inria.fr}%
  \thanks[sfn]{Work done during internship of Rafael Oliveira from}%
  \and
Rafael Oliveira\thanks{Federal University of Pernambuco, rafoli@gmail.com}%
\thanksref{sfn}
}
\authorhead{Deransart \& Oliveira}
\RRtitle{Towards a Generic Framework to Generate Explanatory Traces of Constraint Solving and Rule-Based Reasoning}
\RRetitle{Towards a Generic Framework to Generate Explanatory Traces of Constraint Solving and Rule-Based Reasoning}
\titlehead{Towards a Generic Framework}
\RRresume{
Dans ce rapport, nous montrons comment utiliser le calcul des fluents simple (SFC) pour sp\'ecifier des traceurs g\'en\'eriques, c'est-\`a-dire qui produisent des traces g\'en\'eriques. Une trace g\'en\'erique est une trace qui peut \^etre produite par diff\'erentes impl\'ementations d'un composant logiciel et être utilis\'ees ind\'ependamment du composant trac\'e.

Cette approche est utilis\'ee pour d\'efinir une m\'ethode pour introduire dans une platforme CHR$^\vee$ bas\'ee Java et appel\'ee CHROME (Constraint Handling Rule Online Model-driven Engine) un traceur g\'en\'erique extensible. La m\'ethode comprend une sp\'ecification du traceur en SFC, une m\'ethodologie d'extension, et leur implantation dans CHROME, afin d'obtenir la plateforme CHROME-REF (Raisonnement Explicatif Facilit\'e), qui est un solveur de contraintes et moteur de raisonnement \`a base de r\`egles avec des traces d'explications.
}
\RRabstract{
In this report, we show how to use the Simple Fluent Calculus (SFC) to specify generic tracers, i.e. tracers which produce a generic trace. A generic trace is a trace which can be produced by different implementations of a software component and used independently from the traced component.

This approach is used to define a method for extending a java based CHR$^\vee$ platform called CHROME (Constraint Handling Rule Online Model-driven Engine) with an extensible generic tracer. The method includes a tracer specification in SFC, a methodology to extend it, and the way to integrate it with CHROME, resulting in the platform CHROME-REF (for Reasoning Explanation Facilities), which is a constraint solving and rule based reasoning engine with explanatory traces.
}
\RRmotcle{trace, CHR, CHR$^\vee$, CHROME, CHROME-REF, MDE, traceur, m\'eta-th\'eorie, pilote de tracer, analyseur, outils d'analyse, analyse de programme, analyse dynamique, s\'emantique observationnelle, composant logiciel, d\'eboggage, environnement de programmation, programmation en logique, validation}
\RRkeyword{Trace, CHR, CHR$^\vee$, CHROME, CHROME-REF, Tracer, Meta-Theory, Model Driven Engineering, Tracer Driver, Analysis Tool, Program Analysis, Observational Semantics, Software Component, Debugging, Programming Environment, Logic Programming, Validation}
\RRprojets{Contraintes}
\RRdomaine{2}  
\URRocq 
\RCParis 

\begin{document}
\RRNo{7165}
\makeRR   
\section{Introduction}
\label{intro}

In this report we define a method for extending a java based CHR$^\vee$\footnote{CHR stands for Constraint Handling Rule \cite{fruhwirth2003essentials}, CHR$^\vee$ for CHR with disjunction.} platform called CHROME (Constraint Handling Rule Online Model-driven Engine) with an extensible generic tracer. CHROME is presently developed at the Federal University of Pernambuc \cite{vitorino2009ufpe} with the purpose to allow the development of large software using CHR paradigm.

The method consists of firstly to build a formal specification of a tracer for CHR$^\vee$, the kernel of the system, and to extend it according to further CHR$^\vee$ extensions. The specification defines a generic trace which is as independent as possible from a particular CHR platform or specific usages. Then, secondly, it is suggested to use this specification as a guideline to extend the MDE scheme development of CHROME, with a tracer scheme development, resulting in the plateform CHROME-REF (REF stands for Reasoning Explanation Facilities), which is a constraint solving and rule based reasoning engine with explanatory traces.

This report is almost based on the internship work of R. Oliveira \cite{Rafael09}. We introduce first some contextual aspects of this work.
 
\subsection{The CHR World} 

The language CHR has matured over the last decade to a powerful and
elegant general-purpose language with a wide spectrum of application domains
\cite{sneyrs2003chr}. The interest in CHR$^\vee$ stemmed from past research
having shown that:
  
\begin{itemize}
  \item CHR is simultaneously a Turing-complete declarative programming
  	language and an expressive knowledge representation language with
  	declarative formal semantics in classical first-order logic
  	\cite{fruhwirth2003essentials};
  \item CHR integrates and subsumes the three main rule-based programming and
  	knowledge representation paradigms, i.e., (conditional) term rewrite rules
  	\cite{gamma1995design}, (guarded) production rules \cite{van-efficient} and (constraint)
  	logic programming rules \cite{abdennadher2001rule};
  \item A CHR$^\vee$ inference engine can support an unmatched variety of practical
  	automated reasoning tasks, including constraint solving with variables from
  	arbitrary domains, satisfiability \cite{fruhwirth2003essentials},
  	entailment \cite{schrijvers2006automatic}, 
abduction
  	\cite{abdennadher2001rule}, agent action planning
  	\cite{schrijvers2006automatic} and agent belief update
  	\cite{schrijvers2006automatic} and revision \cite{van-efficient}. In
  	addition, it support several of these tasks under logical \cite{schrijvers2006automatic},
  	plausibilistic \cite{christiansen-prioritized} and probabilistic epistemological
  	\cite{proctor2006drools} assumptions. 
        \cite{SFR08} adds default reasoning to this list,
by showing how to represent default logic theories in CHR$^V$. It also
discusses how to leverage this representation together with the well-know
correspondence between default logic and Negation As Failure (NAF) in
logic programming, to propose an extension CHR$^{V;naf}$ of CHR$^V$ allowing
NAF in the rule heads.
	And \cite{FOM08chr} adds concurrency.
\end{itemize}

\begin{figure} 
\centering 
\scalebox{0.8}{\includegraphics{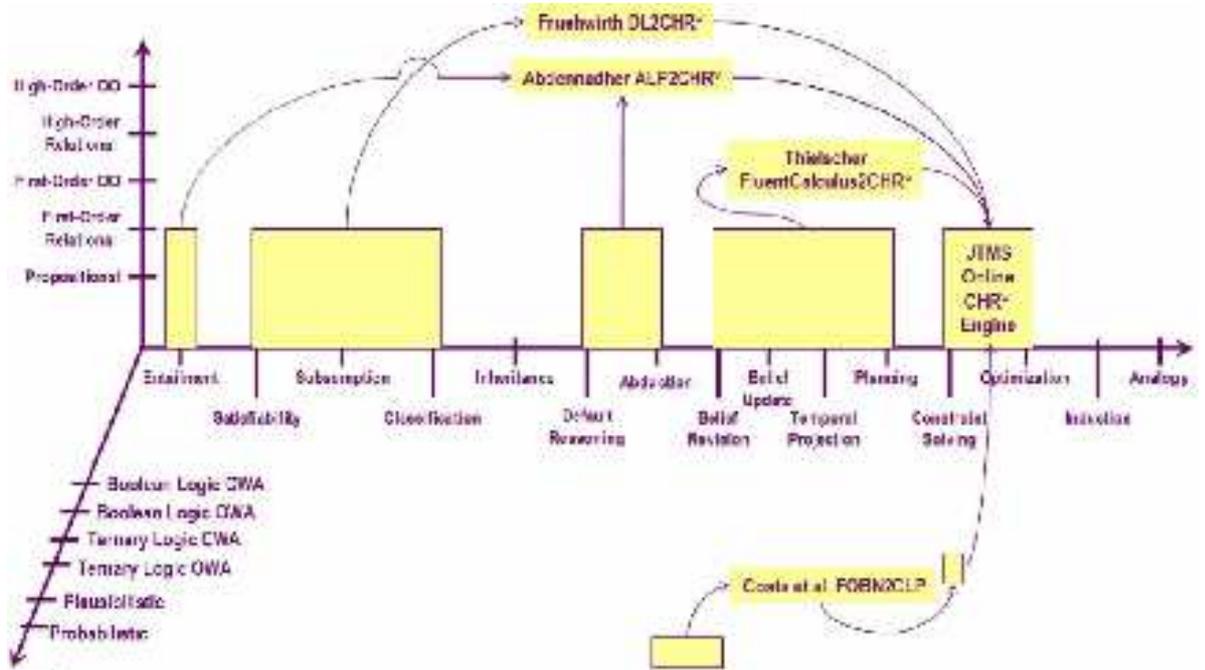}}
\caption[Rule-based constraint programming and AR services]
{Rule-based Constraint Programming and Automated Reasoning Services}
\label{fig:chr}
\end{figure}

The Figure~\ref{fig:chr} shows the several automatic reasoning services that are
subsumed by CHR$^\vee$ and extensions.

For all its above strengths, CHR remained until recently a language for
Knowledge Representation and Programming (KR\&P) in-the-small mainly used to
fast prototype intelligent, innovative systems. The ORCAS
project \footnote{\url{http://www.cin.ufpe.br/~jr/mysite/C4RBCPProject.html}},
constitutes a first and preliminary step towards the long term goal, to turn CHR into a platform for KR\&P in-the-large industrial strength, operationally deployed systems. It addressed one key
KR\&P in-the-large requirement, namely rule base engineering scalability
through rule base encapsulation and assembly in reusable software components. 

\subsection{From CHROME to CHROME-REF}
\label{subs:chref}

CHROME stands for \textbf{Constraint Handling Rule Online Model-driven Engine},
is a model-driven, component-based, scalable, online, Java-hosted CHR$^\vee$
engine to lay at the bottom of the framework as the most widely reused automated reasoning 
component. The idea of CHROME is also to demonstrate how a standard set of
languages and processes prescribed by MDA can be used to design concrete
artefacts, such as: a versatile inference engine for CHR$^\vee$ and its
compiler component that generates from a CHR$^\vee$ base the source code of Java
classes.

\vspace{3mm}
The project CHROME-REF (Constraint Handling Rule Online Model-Driven
Engine with Reasoning Explanation Facilities) constitutes a
preliminary step, towards extending CHR and CHR engines with a formally
founded, flexible and user-friendly reasoning explanation facility. The need
for flexible and user-friendly explanatory reasoning tracing facilities for
rule-based systems has been recognized since the initial success of production
rule expert systems in the 80s. However, the expressive power of CHR being far
superior than that of a mere production system, through the addition of
functional terms, rewrite rules and backtracking search, makes debugging a CHR
rule base also more complex than debugging a production rule base. In turn,
this added complexity makes the need for sophisticated rule engine tracing
facilities more crucial and the issues in their design and implementation more
challenging.

This project is pioneering the investigation of these issues. 
CHROME assembles
CHR base independent run-time components for constraint store management,
fired ruled history management, constraint entailment, query processing and
intelligent search, with optimized components resulting from the compilation
of the CHR base. Following the KobrA2 model-driven, component-based,
orthographic software engineering method \cite{atkinson2000component,robin2009kobra2}, CHROME was built by first specifying
a refined Platform-Independent Model (PIM) in the OMG standards UML2/ OCL2
(Unified Modeling Language / Object Constraint Language). This PIM was then
implemented in Java. 

The fact that CHROME compiling a declarative CHR base
into imperative Java objects is crucial for its reasoning performance. However, it
makes tracing far more complex since it introduces a mismatch between, on the
one hand, the abstract, high-level rule interpretation operational semantics
that the developer follows when conceiving a CHR base, and, on the other hand,
the concrete, low-level object method call operational semantics effectively
executed by the engine. To help the developer debug the rule base, the tracer
must thus generate a high-level rule interpretation trace simulation from the
low-level object method calls executed by the compiled code.

\vspace{3mm}
Our objective here is to integrate the independently constructed tracer architecture
within the component-based architecture of CHROME following the
KobrA2 method.

\subsection{Towards Generic Trace}
\label{sec:chr_project_tmt}

Despite the fact that CHR$^\vee$ provides an elegant general-purpose language with a
wide spectrum of application domains, a key issue is how easily you can write
and maintain programs. Several studies \cite{schneidewind1987state,richer1989evaluation}
\cite{boehm2007spiral} \cite{banker1998software} show that maintenance is the
most expensive phase of software development: the initial development
represents only 20\% of the cost, whereas error fixing and addition of new
features after the first release represent, each, 40\% of the cost. Thus, 80\%
of the cost is due the to the maintenance phase. Debugging is said to be the
least established area in software development: Industrial developers have no
clear ideas about general debugging methods or effective and smart debugging
tools, yet they debug programs anyway. There are
several ways to analyze a program, for instance: program analysis tools help
programmers understand programs, type checkers
\cite{moller-design} help understand data inconsistencies, slicing tools
\cite{korel1997application} help understand dependencies among parts of a
program. Tracers give insights into program
executions.

\begin{figure}
\centering 
\scalebox{0.5}{\includegraphics{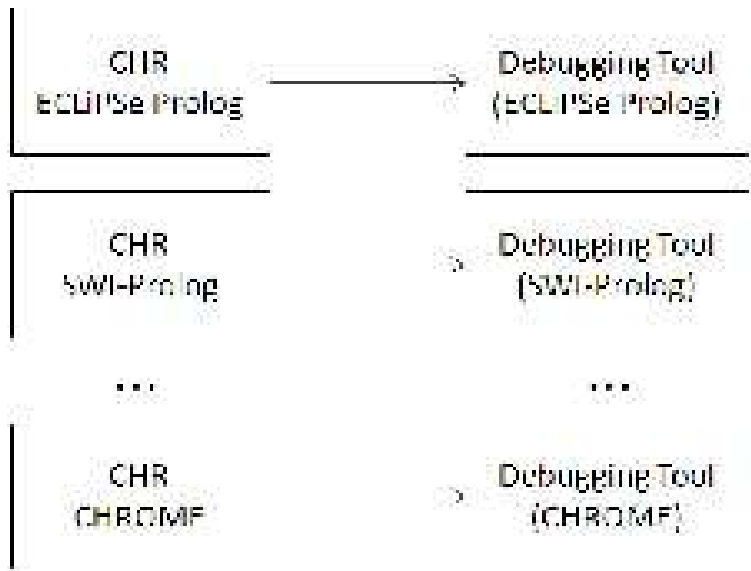}}
\caption[Solver strictly connected to the debugging tool]
{Current situation: each solver is strictly connected to its 
debugging tool. Figure adapted from \cite{langevine2004generic}}
\label{fig:debugging_tools}
\end{figure}

At present, there exists a number of useful debugging tools for CHR, for
example, ECLiPSe Prolog \cite{eclipse-prolog}, SWI-Prolog
\cite{wielemaker2006swi} or CHROME \cite{vitorino2009ufpe}. But, these tools
were designed and implemented in a specific way for each solver, not all tools
benefit from all the existing tool. The Figure~\ref{fig:debugging_tools} shows
this current cenario, for each CHR solver a specific implementation of the
debugging tool.
   
This way each implementation results in a set of one-to-one specialized
connections between a solver and its tools. If we want to interchange
data between each solver, hooks have to be added into the solver code in order
to be able to do it. Furthermore, the types of basic information required by a
given debugging tool is not often made explicit and may have to be
reverse-engineered. This is a non neglectable part of the cost of porting
debugging tool from one constraint programming platform to another.

In order to solve the above-mentioned problem and improve analysis and
maintenance of rule-based constraint programs, like CHR$^\vee$, there is a need for
user-friendly reasoning explanatory facilities that are flexible and portable.

Given this scenario 
we take advantage of the recent research in trace engineering 
\cite{deransart2007observational,deransart2008trace,TMTmanuscript}
to propose a generic architecture that
produces generic debugging informations for CHR$^\vee$ and potential extensions. In that way,
any debugging tool changes its focus to generic traces, instead of to be
concerned in specific platform implementations. 
 
\begin{figure}
\centering
\scalebox{0.9}{\includegraphics{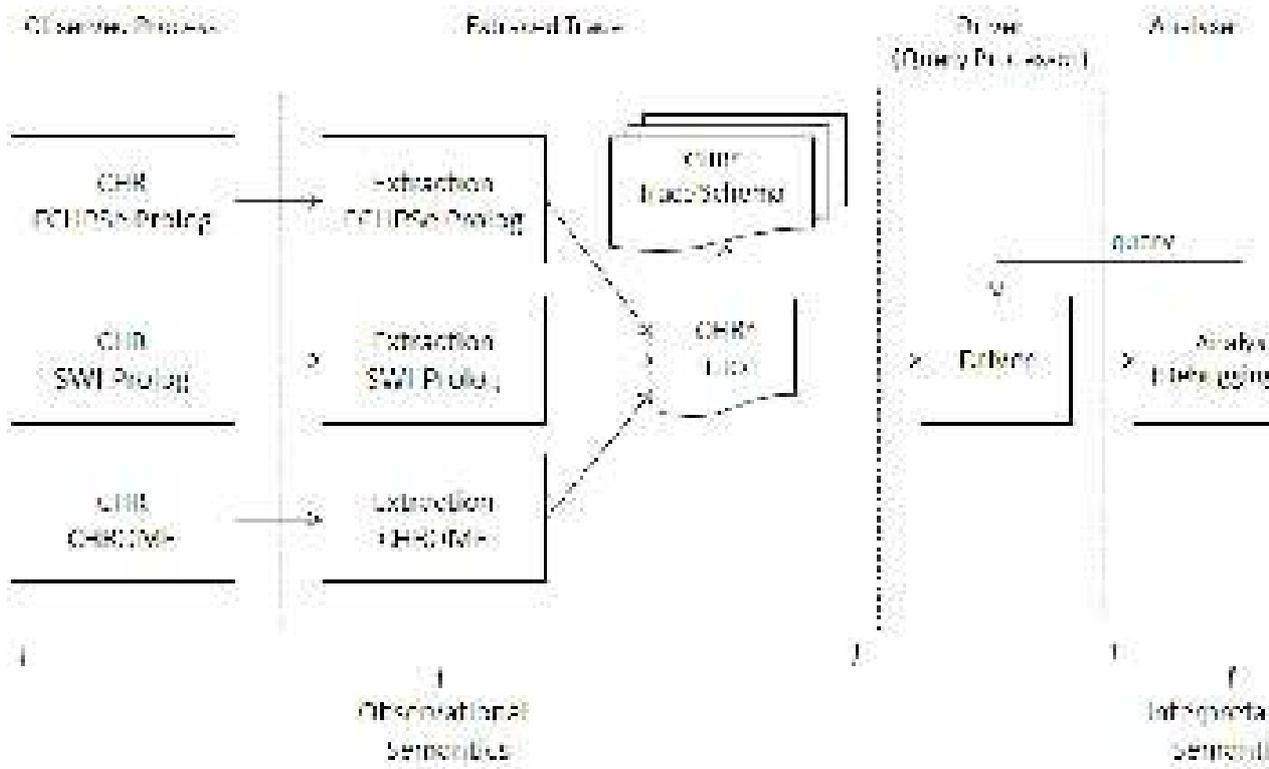}}
\caption[Trace Meta-Theory] {Our approach: a generic trace schema enables work and maintain just one process of debugging}
\label{fig:tmtprocess}
\end{figure}     

The Figure~\ref{fig:tmtprocess} illustrates
the idea of ``generic trace'', which is as independent as possible from a particular CHR platform or specific usages. It shows the structure of a tracing process which can be decomposed into three likely ``independent'' components: trace extraction, full trace filtering according to some query, and reconstruction of a sub-trace to be used. It shows also the several aspects which must be specified: a semantics for the generic trace (called Observational Semantics), a query language to select the sub-trace of interest to be used, and a semantics to interpret the selected trace (called Interpretative Semantics).

\vspace{3mm}
A generic trace needs to be understood independently from the observed process. For this reason it is necessary to be able to give it a semantics as precise as possible. This is the purpose of the Observational Semantics. It will allow for validation tests and studies of some trace properties before and after implementing it. In this report we focus on the observational semantics of traces.

\vspace{1mm}
The Fluent Calculus (FC) is a logic-based representation language
for knowledge about actions, change, and causality \cite{thielscher1999situation}. As an
extension of the classical Situation Calculus \cite{reiter1991frame}, Fluent
Calculus provides a general framework for the development of axiomatic semantics
for dynamic domains. It appears to be well suited to describe the Observational Semantics and, though its Flux implementation \cite{Thielscher98}, to be a likely executable specification.

\subsection{Connecting all the Pieces}

The CHROME project focuses on extending CHR with rule-base encapsulation in software
components for reuse by assembly across applications. 
It has as intention to produce the \textit{first
domain-independent framework highly reusable debugging tool}, supporting a
variety of reasoning explanation facilities. The main contribution is to permit
any CHR$^\vee$ engine to be extensible with components for comprehensive,
flexible and efficient reasoning explanation trace generation and
user-friendly trace query specification and trace visualization. To achieve
this is necessary to integrate design patterns for tracing facilities
such as tracer driver with design patterns for Graphical User Interface
(GUI) such as Model-View-Controller (MVC) within an overall Model-Driven
Architecture (MDA) framework \cite{atkinson2002component}. It will also involve defining a
comprehensive trace query language, as well as experiments to empirically
evaluate the engineering productivity gains obtained through the use of the
tracing components.

\begin{figure}
\centering
\scalebox{0.9}{\includegraphics{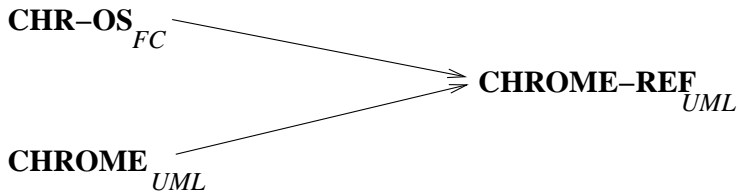}}
\caption[A general approach to trace CHROME] {Towards CHROME-REF}
\label{fig:general_approach_trace}
\end{figure}
 
\vspace{2mm} 
In the report we describe the approach illustrated by the Figure~\ref{fig:general_approach_trace}.
It consists first in an observational semantics of the extensible generic trace 
of CHR which is specified in Fluent Calculus.
This semantics is thus mapped into the PIM description of CHROME, leading to a complete PIM of CHROME-REF in UML, the constraint solving and rule based reasoning engine with explanatory traces.

The CHROME-REF environment will be built such an editor as a Eclipse Plugin
\footnotetext[\value{footnote}]{http://www.eclipse.org/} for rapid prototyping
deployed with a GUI to interactively submit requests and inspect solution
explanations at various levels of details.


\vspace{5mm}
The rest of this report is organized in three main sections.

The Section~\ref{sec:TraceMetaTheory} presents a restricted trace meta-theory focused on trace production components and composition. It introduces also the observational semantics of a trace and its representation in the simplified fluent calculus.

The Section~\ref{sec:rulebasedsystems} presents the observational semantics of CHR$^\vee$ in fluent calculus including tracer and extraction schemes.

The Section~\ref{chromeref} shows the introduction of the tracer in the PIM of CHROME using the KobrA2 method and resulting in a PIM of CHROME-REF with a very first implementation.

Four annexes give respectively a description of the Observational Semantics of CHR in SFC, the XML scheme of a generic trace of CHR$^\vee$, a short example of trace produced by the java compiled CHROME-REF, and the OS of an application.


\newpage
\section{Specifying Tracers}
\label{spectrac}
\label{sec:TraceMetaTheory}

The Trace Meta-Theory (TMT) \cite{TMTmanuscript} provides a set of definitions about how to design a trace for a specific domain of observation.

A \textbf{trace} may be interpreted as a sequence of communication actions
that may take place between an \textbf{observer} and an
\textbf{observed process}. It consists of
finite unbounded sequences trace events. There
is also the \textbf{tracer} that means the generator of trace. According
to \cite{deransart2008trace}, the TMT focuses particularly on providing semantics to 
tracers and the produced traces as independent as
possible from those of the processes or from the ways the tracers produce them.

There are two concepts of trace \cite{deransart2008trace} (cf. the Figure~\ref{fig:chromepipiline} and the Section~\ref{sec:conttrace}). The first one is the \textbf{virtual trace}, it
represents a sequence of events showing the evolution of a virtual state which contains
all that one can or wants to know about the observed process. The second one
is called \textbf{actual trace}, it represents the generated trace in the form of some encoding of the current virtual state. Finally, there is the idea of \textbf{full trace} if the parameters chosen to be observed about the process
represent the totality of useful knowledge regarding it (explicitly or implicitly).

\subsection{Components of Trace Generation and Use}

\begin{figure}  
\centering
\scalebox{0.6}{\includegraphics{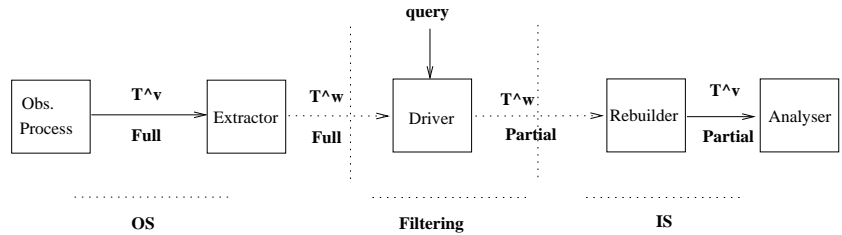}}
\caption[Components of the TMT, Composants pour la production et l'utilisation d'une trace]
{Components of the TMT}
\label{fig:componentsPD}  
\end{figure}

The Figure~\ref{fig:componentsPD} shows the different components related to a unique trace.
We distinguish  5 components, in this order.
\begin{enumerate}
\item Observed process

The observed process is assumed more or less abstracted in such a way that his behavior can be described by a virtual trace, that is to say, a sequence of (partial) states. A formal description of the process, if possible, can be considered as a formal semantics, which can be used to describe the actual trace extraction.

\begin{figure} [h]
\centering
\scalebox{0.3}{\includegraphics{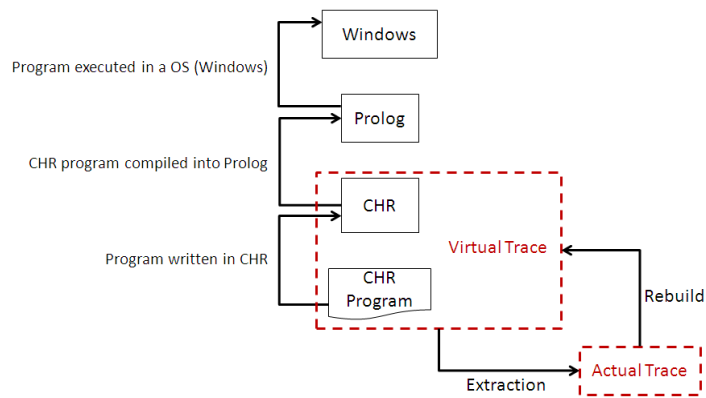}}
\caption[Virtual and Actual Trace]
{Virtual and Actual Trace.}
\label{fig:chromepipiline}
\end{figure}

\item Extractor

This component is the extraction function of the actual trace from the virtual trace. From a theoretical point of view, we can see it as a specific component, but in practice it corresponds to the tracer whose realization, in the case of a programming language, usually requires modifying the code of the process.
\item Filter

The role of the filter component, or {\em driver} \cite{DLads05}, is to select a useful sub-trace. This component requires a specific study. It is assumed here that it operates on the actual trace (that produced by the tracer). The fact of making it as a proper component corresponds to the specific approach adopted here, which implies that the extracted actual trace is full. The filtering depends on the specific application, implying that the full trace already contains all the information potentially needed for various uses.
\item Rebuilder

The reconstruction component performs the reverse operation of the extraction, at least for a subpart of the trace, and then reconstructs a sequence of partial virtual states. If the trace is faithful (i.e. no information is lost by the driver) \cite{TMTmanuscript}, this ensures that the virtual trace reconstruction is possible. In this case also, the separation between two components (rebuilder and analyzer) is essentially theoretical; these two components may be in practice very entangled.
\item Analyzer

The component using a trace may be a trace analyzer or any application. 
\end{enumerate}

\begin{figure} 
\centering
\scalebox{0.7}{\includegraphics{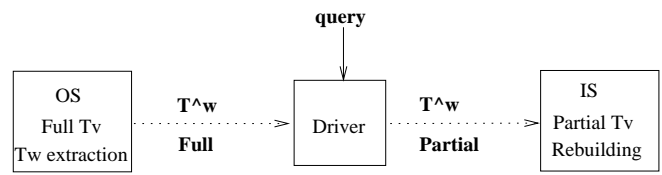}}
\caption[Concepts formels li\'es \`a la production et l'utilisation d'une trace]
{Formal concepts related to the generation and use of a trace}
\label{fig:componentsso}  
\end{figure}

\vspace{4mm}
With these components it may be associated three main specification steps, as illustrated on the Figure~\ref{fig:componentsso}.

\begin {enumerate}[-]
\item Observational Semantics (OS)

The OS describes formally the observed process (or a family of processes) and the actual trace extraction. This aspect will be studied deeper in the Section~\ref{sec:conttrace}. The intention here is to express the OS using simple fluent calculus.
\item Querying

Due to the separation in several components, the actual trace may be expressed in any language. We suggest using XML. This allows to use standard querying techniques defined for XML. This aspect will not be developed here, but we chose to express the trace in XML and give in the Appendix~\ref{app:chrtraceschema} the corresponding XML schema.
\item Interpretative Semantics (IS)

The interpretation of a trace, i.e. the capacity of reconstructing the sequence of virtual states from an actual trace, is formally described by the Interpretative Semantics. In the TMT no particular application is defined; its objective is just to make sure that the original observed semantics of the process has been fully communicated to the application, independently of what the application does.
\end{enumerate}

\subsection{Contiguous Full Traces}
\label{sec:conttrace}

We introduce here the two traces which may be associated to a single process equipped with a tracer. We recall here the definitions used in \cite{TMTmanuscript}.


\subsubsection{Virtual Trace}

A full virtual trace is defined on a domain of states. Given ${\cal P}$  a finite set of (names) of parameters $ p_i $  defined on the domains ${\cal P}_i$.
The ${\cal P}_i$ are domains of objects of any kind. They may also have relations (functional or otherwise) between them and they can be infinite in size.

A domain of states ${\cal S}$  is defined on the Cartesian product of the parameter domains: $S \subseteq {\cal P}_{1}\times ...\times {\cal P}_{n}$.

\begin{definition} [Contiguous Full Virtual Trace]

A contiguous full virtual trace  is a sequence of trace events of the form {\bf $e_t: (t, r_t, s_t),\ t \ge 1$}, where:
\begin{itemize}
  \item $t$: is the {\bf chrono}, specific time of the trace. It is an integer increased by one unit in each successive event. To point a particular value of the chrono, we will talk about {\em moment} of the trace.
   \item $r_t$: an identifier of {\bf action} characterizing the type of actions undertaken to make the transition from state $s_{t-1}$ to state $s_t$.
   \item $s_t$: is an element of the state domain. $s_t = p_{1, t}, ..., p_{n, t}$ is the current state reached
at moment $t$, and the $p_{i, t}$ are values of the {\bf parameters} $p_i$ at moment $t$.
$s_t$ is the {\em current full virtual state}.
\end{itemize}
\end{definition}
A finite virtual trace over $t\ (t > 0)$ events will be denoted $T^v_t = <s_0, \overline{e_t}>$, where $s_0$ is the initial full virtual state and $\overline{e_t}$ represents the sequence $e_1 \ldots e_i \ldots e_t $.


The full virtual trace is {\em contiguous} insofar as all the moments in the interval $[1..t]$ are present in the trace $T^v_t = <s_0, \overline{e_t}>$.

\subsubsection{Actual Trace}

The full virtual trace represents what we want or what is possible to observe of a process. It describes the development stages of this process in the form of the evolution of a state which contains the observables. As the current virtual state of a process can be fully represented in this trace, one cannot expect neither to produce it nor to communicate it efficiently. In practice we will perform a kind of ``compression'' of the information conveyed by the virtual states and their evolution, transmitted or communicable to the process observers, and one shall ensure that these processes are able to ``decompress'' it. This actually communicated information is the {\em actual trace}.

An actual full trace is defined on an actual state domain. Let ${\cal A}$ be a finite set of (names of) attributes $a_i$ defined on domains of attributes ${\cal A}_i$.
The attributes may have relationships (they are not necessarily independent) and they can be infinite in size.

An actual state domain ${\cal A}$ is defined on the Cartesian product of attributes domains: ${\cal A} \subseteq {\cal A}_{1}\times ...\times {\cal A}_{n}$. 

\begin{definition} [Contiguous Full Actual Trace]

An actual trace is a sequence of trace events of the form {\bf $w_t: (t, a_t),\ t \geq 1$}, where:

\noindent
$t$ is the chrono and $a_t\ \in\ {\cal A}$ denotes a finite sequence of attributes values. $a_t$ is the current actual state. The number of attributes of a trace event is bound by $ n $. Each state $a_t$ contains at most $n$ attributes whose number depends exclusively on the type of action which produced it.
\end{definition}
An actual trace with $t\ (t > 0)$  events is denoted $T^w_t = <s_0, \overline{w_t}>$, where $s_0$ is the initial virtual state common to both traces and $\overline{w_t}$  represents the sequence $w_1, \ldots w_i, \ldots, w_t $.


\subsection{Generic Trace and Composition}

We study here the methodology of generic full trace development for a multi-layer based application.

\subsubsection{Generic Trace of a Familly of Observed Processes}

Consider again the Figure~\ref{fig:tmtprocess} in the introduction.
It illustrates the fact that different implementations of CHR can be abstracted by a unique simpler model.
This common model is used to specify the unique virtual and actual traces of these implementations.
This illustrates the way we will proceed to get a generic trace of CHR: starting from an abstract theoretical, general but sufficiently refined, semantics of CHR which is (almost) the same implemented in all CHR platforms.

\subsubsection{Composition of Traces}
\label{sec:compos}

Now we consider the case of an application written in CHR. It may be for example a particular constraints solver like CLP(FD). In this case there exists already a generic trace called {\em GenTra4CP} \cite{oadimpac}. This trace is generic for most of the CLP(FD) existing constraints solvers. Therefore a tracer of CLP(FD) solver implemented in CHR should also produce this trace. But we may be interested in refining the trace considering that there are two layers: the layer of the application (CLP(FD)) and the layer of the language in which it is implemented (CHR). The most refined trace will then be the trace in the GenTra4CP format extended with elements of the generic full trace of CHR alone. The generic full trace of CLP(FD) on CHR is an extension of the application trace taking into account details of lower layers. 

\begin{figure} 
\centering
\scalebox{0.6}{\includegraphics{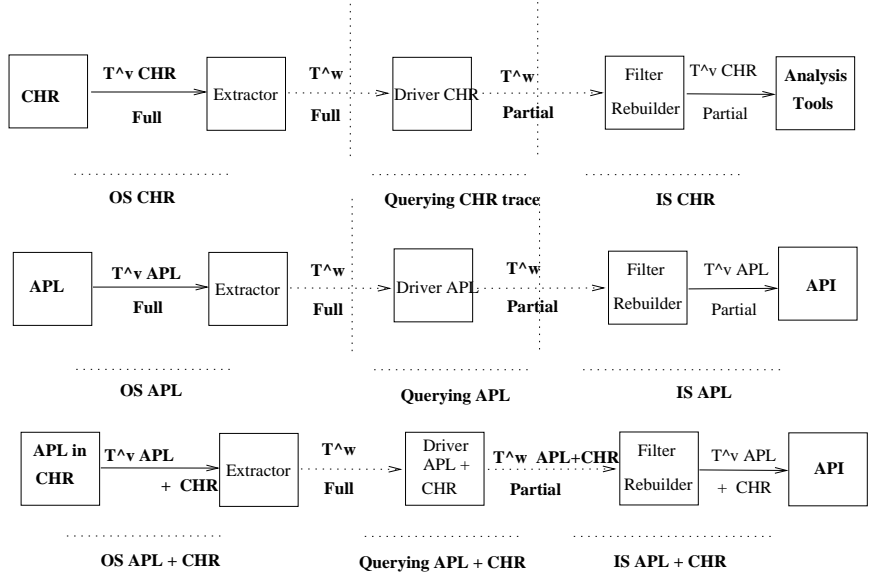}}
\caption[Combination of Generic Full Traces for a two Layers Application]
{Composition of Generic Full Traces for a two Layers Application}
\label{fig:componentsaplchr}
\end{figure}

This is illustrated by the Figure~\ref{fig:componentsaplchr} in the case of two layers: an application (like CLP(FD) for example) implemented in CHR. This method can be generalized to applications with several layers of software. The Figure~\ref{fig:chromepipiline} shows in fact at least 4 layers.

In our components based approach it means that we may define separately and independently specific generic full traces for each layer, and, so in this case for the application (APL) and the under-layer of CHR. The generic full trace APL on CHR is a kind of {\em composition} of traces and will be obtained by some merging of both generic full traces into a unique one. The result may not exactly be a union of all actions, parameters and attributes, but it is not our purpose here to study more deeply this aspect. For more details see \cite{TMTmanuscript}.



\subsection{Observational Semantics}

The Observational Semantics (OS) is a description of a possibly unbounded
data flow without explicit reference to the operational semantics of the
process which produced it \cite{deransart2007observational}. The
OS may be considered as a abstract model of process, in the
case of a single observed process or it can be an abstraction of the
semantics to several processes. It is defined as a Labelled
Transition Systems (LTS) \cite{gilmore2005trends}.


%

\subsubsection{Representation of the Observational Semantics}

The Observational Semantics has two parts: a state transition function and a trace extraction function.

The first part is a formal model on the way successive events of the virtual trace are related. 
It is a virtual trace semantics in the sense that, given a full virtual trace \newline $T^v_t = <s_0, \overline{e_t}>$, it explains the sequence of events $e_t$ by a transition function\footnote{It is fact a relation since the transitions may be nondeterministic.} recursively applied from an initial virtual state.


\vspace{1mm}
The second part, the function of extraction, produces what is actually ``broadcasted'' outside from the observed process. This function has as arguments the current state and the type of action, and produces the attributes of the actual trace.

\begin{definition} [Observational Semantics (OS)]

An Observational Semantics is defined by the tuple $< S, R_O, A, E, T, S_0 >$, where
\begin{itemize}
\item $S$ is a  {\em virtual state domain}, where each state is described by a set of parameters.
\item $R_O$ is a finite set of action types, set of identifiers used as labels for transitions.
\item $A$ is a {\em actual state domain}, where each state is described by a set of attributes.
\item $E$ is the local extraction function of the actual state $a$, performed by transition of action type $r$ issued from state $s$, $E : R$ {\tt x} $S \rightarrow A$, which satisfies by definition: $E(r,s)= a$ ($a \in A$, set of actual states). More precisely, the set of attributes $a_t$ of the event $t$ of the actual trace is derived from the current state at moment $t-1$ of the virtual trace and the transition labelled by the action type $r_t$, i.e.
\begin{quote}
$E(r_t, s_{t-1})= a_t$
\end{quote}
\item $T$ state transition function $T : R$ {\tt x} $\rightarrow S$, i.e.
\begin{quote}
$T(r_t, s_{t-1})= s_t$
\end{quote}
\item $S_0 \subseteq S$, set of initial states.
\end{itemize}
\end{definition}

\vspace{2mm}

The OS may be represented by ``rules'', one for each action, describing the transition and the actual trace event extraction corresponding to the action. A rule has 4 items.

\vspace{0,5mm}
\begin{itemize}
\item {\bf AType}: an action identifier $r \in R_O$
\item {\bf ACond}: \{ some auxiliary computations on the current virtual state and condition for executing the action corresponding to the transition: a first-order logic formula using predicates on the parameters \}
\item {\bf VSEffect}: \{ the effect of the action $r$ on the current state $s$, resulting in a new state $s'$ , and some auxiliary computations relative to the attributes of the trace event \}
\item {\bf Etrace}: \{ the attributes of the trace event produced by the action $r$: $a$, new extracted actual trace event \} 
\end{itemize}


\vspace{5mm}
{\bf Example 2.1: the Fibonacci Function}
\label{ex:fibo}

Idealized (biologically unrealistic) rabbit population.

\vspace{2mm}
The OS $< S, I_f, R_O, A, E, T, S_0 >$, describes the deterministic transition function $T_l$.

\vspace{2mm}
$S$: ${\cal N}_+^*$ (positive integers list), $s_t$ is the complete evolution of the population from moment $0$ until moment $t+1$: $s_t = [popu_0,\ldots, popu_t, popu_{t+1}]$ 

$R_O$: \{$mg$\} (monthly growing)  

$A$: ${\cal N}_+$, $a_t$ is the population at moment $t+1$ ($popu(t+1)$).

$E$: $E(mg,s) = plast(s)+last(s)$. There is one rule only to describe $E$.

$T_l$: $T(mg,s) = s \ o \ [plast(s)+last(s)]$ (respectively before last and last elements of the list s$s$, $o$ denote lists concatenation). The new virtual state $t$ is the previous state to which the sum of the two last elements is appended.

$S_0$: $s_0 = [1,1]$.


\vspace{1mm}
{\bf AType}: $mg$

{\bf ACond}: \{ $true$\}

{\bf VSEffect}:  \{ $v \leftarrow plast(s)+last(s) \wedge \ s' \leftarrow s\  o \  [v]$ \}

{\bf Etrace}:  \{$v$\} 

\vspace{4mm}
{\bf Traces:}
\ \ 
\newline $T^v_5 = <[1,1],[(1,mg,[1,1,2]), (2,mg,[1,1,2,3]), \ldots,$

$ \ \ \ \ \ \ \ \ \ \ \ \ \ \ \ \ \ \ \ \ \ \ \ \  (4,mg,[1,1,2,3,5,8]),(5,mg,[1,1,2,3,5,8,13])]>$
\newline $T^w_5 = <[1,1],[(1,2),(2,3),(3,5),(4,8),(5,13)]> $

\subsubsection{Simple Fluent Calculus}\label{sec:simplefluentcalculus}
\label{FC}

The Fluent Calculus (FC) is a logic-based representation language
for knowledge about actions, change, and causality \cite{thielscher1999situation}. As an
extension of the classical Situation Calculus \cite{reiter1991frame}, Fluent
Calculus provides a general framework for the development of axiomatic semantics
for dynamic domains.

\vspace{1mm}
The simple fluent calculus (SFC) has the following appealing qualities:
\begin{enumerate} [-]
\item Simplification of the description, since the notions of virtual state and actual trace are ``naturally'' embedded in the fluent calculus (the situation corresponding to a state can be viewed as a representation of the actual trace).

\item Reasoning on transitions may be simpler, as it supports handling partial virtual states (with appropriate axiomatisation). Formal proofs become simpler in the SFC  (less deductions, at least for ``direct closed'' effects, see \cite{thielscher1999situation,framepb03}). It follows that properties like ``faithfulness'' \cite{deransart2007observational} should be easier to prove formally. The Symmetry of ``state axioms'' allows forward and backward reasoning. With the simplicity of the representation of state changes, this should make such proofs simpler.

\item The description in SFC makes such specification potentially executable in Flux. There is some limits to the executability, related to the partial axiomatisation of the observational semantics and executability of formal specification in general. However such framework may facilitate some simulations.

\end{enumerate}

The Fluent Calculus is a sorted logic language with four standard sorts: FLUENT,
STATE, ACTION, and SIT (which stands for situation). A fluent describes a single state property
that may change by the means of the actions of some agent. A state is a collection of fluents.
Adopted from Situation Calculus, the standard sort SIT describes sequences of actions.

The pre-defined constant $\emptyset : STATE$ stands for the empty state. Each
term of sort FLUENT is also an (atomic) STATE, and the function $\circ : STATE
\times STATE \mapsto STATE$, written in infix notation, represents the composition
of two states. The following abbreviation $Holds(f, z)$ is used to express that
fluent $f$ holds in state $z$:
\begin{equation}
Holds(f, z) =_{def} (\exists z') f \circ z' = z
\end{equation}

The behavior of function ``$\circ$'' is governed by the
foundational axioms of Fluent Calculus, which essentially characterize states as sets of fluents.

\begin{equation}
( z_1 \circ z_2 ) \circ z_3 = z_1 \circ ( z_2 \circ z_3 )
\end{equation}

\begin{equation}
z_1 \circ z_2 = z_2 \circ z_1
\end{equation}

\begin{equation}
z \circ z = z
\end{equation}

\begin{equation}
z_1 \circ \emptyset = z
\end{equation}

\begin{equation}
Holds(f, f_1 \circ z) \supset f_1 \vee Holds(f, z)
\end{equation}

States can be updated by adding and/or removing one or more fluents. Addition of
a sub-state $z$ to a state $z_1$ is simply expressed as $z_2 = z_1 \circ z$, and
removal is defined by

\begin{equation}
z_2 = z_1 - z =_{def} (Holds(f, z_2) \equiv Holds(f,
z_1) \land \neg Holds(f, z))
\end{equation} 

The standard predicate $Poss : ACTION \times STATE$ in Fluent Calculus is used to
axiomatize the conditions under which an action is possible in a state, i.e., the situations
in which the \textit{pre-condition} of this actions is satisfied.

The pre-defined constant $S_0 : SIT$ is the initial (i.e., before the execution
of any action) situation. The function $Do : ACTION \times SIT \mapsto SIT$ 
denotes the addition of an action to a situation. The standard function $State
: SIT \mapsto STATE$ is used to denote the state, i.e., the fluents that hold in
a situation, after a sequence of actions. This allows to extend macro Holds and 
predicate $Poss$ to SITUATION arguments as follows.

\begin{equation}
Holds(f, s) =_{def} Holds(f, State(s))
\end{equation}  
\begin{equation}
Poss(a, s) =_{def} Poss(a, State(s))
\end{equation} 

In a Fluent Calculus Axiomatization, beyond the definition of the domain sorts, functions
and predicates, we can define a set of axioms that must follow three pre-defined
axiom schemas: the precondition axioms, the state update axioms and the state constraint
axioms.
 
\begin{definition}[Pure State Formula]
A Pure State Formula is a First Order formula
$\Pi(z)$

\begin{itemize}
    \item There is only one free state variable $z$
	\item It is composed of atomic formulas in the form:
	\subitem $Holds(\phi, z)$, where $\phi$ is of the sort FLUENT
	\subitem atoms which do not use any reserved predicate of Fluent Calculus 
\end{itemize}

\end{definition}

\begin{definition}[Precondition Axiom]
A precondition axiom follows the schema: $Poss(A(\vec{x}), z) \equiv
\Pi(\vec{x}, z)$, where $\Pi(\vec{x}, z)$ is a Pure State Formula.
\end{definition}

This kind of axiom states that the execution of the action $A$ with the
parameters $\vec{x}$ is possible in the state $z$ if and only if $\Pi_A(\vec{x},
z)$ is true.

%
%

\begin{definition}[State Update Axiom]
A state update axiom follows the schema:

$Poss(A(\vec{x}), State(s)) \land \Pi(\vec{x}, State(s)) \supset \Gamma(State(Do(A(\vec{x}), s)), State(s))$

where

$\Gamma(State(Do(A(\vec{x}), s)), State(s)) = State(Do(A(\vec{x}), s)) = State(s) \circ \vartheta^{+} - \vartheta^{-}$

where

$\vartheta^{+}$ and $\vartheta^{-}$ are partial states.
\end{definition}

\subsubsection{Observational Semantics in Simple Fluent Calculus}
\label{sec:obssemSfc}

\vspace{2mm}
A virtual state of the observed process corresponds to a state in SFC described by a set of fluents (this correspondence must be explicitly specified).

Each type of action in the OS is an action name in the SFC. A particular action is denoted $R$ in the following.

Actual states are elements of the Cartesian product of attribute domains (this domain must be explicitly specified).

Transition and extraction function (or relation) are described using both fundamental following schemes (fundamental axioms of the Fluent calculus)

\begin{enumerate}
\item \textit{Pre-Condition Axioms}:
 
$\ \ \ Poss(R, \vec{x}, z)  \equiv \Pi(\vec{x}, z)$


\item \textit{State Update Axioms}: 

$\ \ \ Poss(R, \vec{x}, State(s)) \land \Pi(\vec{x}.\vec{y}, State(s)) \supset $

$\ \ \ \ \ \ \ \ \ \ \ \ \ \ \ \ \ \Gamma_R(State(Do(R, w(\vec{x}.\vec{y}, State(s)),s)), State(s))$

where $w(\vec{x}.\vec{y},State(s))$ is an actual trace event associated with the transition, and derived from the current state (using local variables too).
\end{enumerate}

There are as many pre-conditions and state update axioms as there are action types $R$ in the OS. $\Gamma_R$ may be a disjunction. It defines the new virtual state and the corresponding extracted actual trace event attributes $w(\vec{x}.\vec{y},State(s))$.

\vspace{1mm}
Nota: a situation $s$ contains the sequence of actions in the OS executed to reach the current virtual state $z=State(s)$, and also the sequence of extracted actual trace events such that $T^w = E(T^v)$.

\vspace{1mm}
An actual trace $T^w$ is the sequence of $w_i$, with chrono, found in the situation 

$\ \ \ s = Do(R_n, \vec{x}_n, w_n, Do(R_{n-1}, \vec{x}_{n-1}, w_{n-1}, ..., S_0)...)$

It may be computed according to the following axioms:

$\ \ \ Extraction(0,S) = S$

$\ \ \ Extraction(n+1, Do(R,\vec{x}, w, s)) = ((n+1).w).Extraction(n, s)$

\vspace{4mm}
{\bf Example 2.2: OS for Fibonacci}
\label{ssec:fibSFC}

The virtual state contains only one fluent $Fib/1$ of type $List(Int) -> Fluent$, and there is only one type of action $Mg$. The actual state contains just 2 attributes, respectively of type $String$ and $Int$. A vector is represented by a sequence (Prolog list syntax).

%




\vspace{4mm}

$
S_0 = Fib([1,1])
$ \\

$
Poss(Mg, [l, pl], z)  \equiv Holds(Fib([l,pl|x]), z) 
$ \\

$
Poss(Mg, [l, pl], State(s)) \land v = l+pl \supset $

$\ \ \ \ State(Do(Mg, [Mg, v], s)) = State(s) \circ Fib([v,l,pl|x]) - Fib([l,pl|x])
$

\subsection{Trace Query and Analysis Tools}



Trace query and analysis tools are considered here as separate components, as illustrated in the Figure~\ref{fig:componentsPD}. Although they are part of the CHROME-REF project, they are not studied more deeply here. Instead, we propose a first repesentation of the actual CHR trace using XML. The XML schema is given in the Appendix~\ref{app:chrtraceschema}, and an example of produced trace in the XML format in the Appendix~\ref{app:leqexample}.


\newpage
\section{Tracing Rule-Based Constraint Programming}
\label{rulecp}
\label{sec:rulebasedsystems}


Constraint Handling Rules emerges in the context of Constraint
Logic Programming (CLP) as a language for describing Constraint Solvers. In
CLP, a problem is stated as a set of constraints, a set of predicates and a
set of logical rules. Problems in CLP are generally solved by the interaction
of a logical inference engine and constraint solving components. The logical
rules (written in a host language) are interpreted by the logical inference
engine and the constraint solving tasks are delegated to the constraint
solvers.

\subsection{CHR by Example}
\label{chr}

The following rule base handles the less-than-or-equal problem: 
\lstset{
	backgroundcolor==\color{darkgray},
	tabsize=4,
	rulecolor=,
	language=Prolog,
        basicstyle=\footnotesize,  
        upquote=true,
        aboveskip={1.5\baselineskip},
        columns=fixed,
        showstringspaces=false,
        extendedchars=true,
        breaklines=true,
        prebreak = \raisebox{0ex}[0ex][0ex]{\ensuremath{\hookleftarrow}},
        showtabs=false,
	numberstyle=\tiny,
	numbersep=5pt,
    showspaces=false,	
    showstringspaces=false,
    identifierstyle=\ttfamily,
    keywordstyle=\color{black}\bfseries\emph,
    commentstyle=\color[rgb]{0.133,0.545,0.133},
    stringstyle=\color[rgb]{0.627,0.126,0.941},
}

\begin{lstlisting}
reflexivity   r1@ leq(X,Y) <=> X=Y | true.
antisymmetry  r2@ leq(X,Y) , leq(Y,X) <=> X=Y.
idempotence   r3@ leq(X,Y) \ leq(X,Y) <=> true.
transitivity  r4@ leq(X,Y) , leq(Y,Z) <=> leq(X,Z).
\end{lstlisting}

\vspace{1mm}
This CHR program specifies how $leq$ simplifies and propagates as a constraint.
The rules implement reflexivity, antisymmetry, idempotence and transitivity in a
straightforward way. CHR $reflexivity$ states that $leq(X,Y)$ simplifies to
$true$, provided it is the case that $X=Y$. This test forms the (optional)
guard of a rule, a precondition on the applicability of the rule. Hence,
whenever we see a constraint of the form $leq(X,X)$ we can simplify it to true.

\vspace{2mm}
The rule $antisymmetry$ means that if we find $leq(X,Y)$ as well as $leq(Y,X)$
in the constraint store, we can replace it by the logically equivalent $X=Y$.
Note the different use of $X=Y$ in the two rules: in the $reflexivity$ rule the
equality is a precondition (test) on the rule, while in the $antisymmetry$ rule
it is enforced when the rule fires. (The reflexivity rule could also have been
written as $reflexivity @ leq(X,X) <=> true.$) 

\vspace{2mm}
The rules $reflexivity$ and $antisymmetry$ are \textit{simplification CHR}. In
such rules, the constraint found are removed when the rule applies and fires.
The rule $idempotence$ is a \textit{simpagation CHR}, only the constraint in the right part
of the head will be removed. The rule says that if we find $leq(X,Y)$ and another
$leq(X,Y)$ in the constraint store, we can remove one.

\vspace{2mm}
Finally, the rule $transitivity$ states that the conjunction $leq(X,Y), leq(Y,Z)$
implies $leq(X,Z)$. Operationally, we add $leq(X,Z)$ as (redundant) constraint.
without removing the constraints $leq(X,Y), leq(Y,Z)$. This kind of CHR is
called \textit{propagation CHR}.

\vspace{3mm}
The CHR rules are interpreted by a CHR inference engine by rewriting the initial
set of constraints by the iterative application of the rules. Its extension with disjunctive
bodies, CHR$^\vee$ boosts its expressiveness power, turning it into a general
programming language (with no need of an host language).

\subsection{Observational Semantics of CHR}
\label{sec:omegat}

The observational semantics of a tracer is based on a simplified abstract semantics of the observed process. In the case of CHR$^\vee$, we suggest to use an adaptation of the refined theoretical semantics of CHR as presented in \cite{duck2004refined}. To start with, we show how to build an observational semantics for CHR based on the theoretical operational semantics $\omega_t$ \cite{fruhwirth2003essentials}.
The description of $\omega_t$ in SFC is borrowed from \cite{silva2009ufpe}.

\subsubsection{Theoretical Operational Semantics $\omega_t$ of CHR}
\label{sec:obsemChrSfc}

We define $CT$ as the constraint theory which
defines the semantic of the built-in constraints and thus models the internal
solver which is in charge of handling them. We assume it supports at least the
equality built-in. We use $[H|T]$ to indicate the first ($H$) and the remaining
($T$) terms in a list, $++$ for sequence concatenation and $[]$ for empty
sequences.

\vspace{1mm}
We use the notation ${a_0, \ldots , a_n}$ for both bags and sets. Bags are sets
which allow repeats. We use $\cup$ for set union and $\uplus$ for bag union,
and \{\} to represent both the empty bag and the empty set. The identified
constraints have the form $c\#i$, where c is a user-defined constraint and i a
natural number. They differentiate among copies of the same constraint in a
bag. We also assume the functions $chr(c\#i) = c$ and $id(c\#i) = i$.

\vspace{2mm}
An execution state is a tuple $\langle Q, U, B, P \rangle_n$, where $Q$ is the
Goal, a bag of constraints to be executed; $U$ is the UDCS (User Defined Constraint
Store), a bag of identified user defined constraints; $B$ is the BICS (Built-In
Constraint Store), a conjunction of constraints; $P$ is the Propagation History,
a set of sequences, each recording the identities of the user-defined
constraints which fired a rule; $n$ is the next free natural used to number an
identified constraint.

\vspace{2mm}
The initial state is represented by the tuple $\langle Q, [], true, []
\rangle_n$. The transitions are applied non-deterministically until no
transition is applicable or the current built-in constraint store is
inconsistent. These transitions are defined as follows:

\vspace{3mm}
\begin{center}
\fbox{\begin{tabular}{p{3.5in}}        
\textbf{Solve} $\langle \{c\} \uplus Q, U, B, P \rangle_n \mapsto \langle Q, U,
c \wedge B, P \rangle_n$ where $c$ is built-in
\newline
\textbf{Introduce} $\langle \{c\} \uplus Q, U, B, P \rangle_n \mapsto \langle Q,
\{c\#n\} \uplus U, B, P \rangle_{n+1}$ where $c$ is user-defined constraint
\newline
\textbf{Apply} $\langle Q, H_1 \uplus H_2 \uplus U, B, P \rangle_n \mapsto
\langle C \uplus Q, H_1 \uplus U, e \wedge B, P' \rangle_n$ where exists a rule
$r @ H'_1 \setminus H'_2 \Leftrightarrow g \lfloor C$ and a matching substitution $e$,
such that $chr(H_1) = e(H'_1), chr(H_2) = e(H'_2)$ and $CT \models B \supseteq
\exists(e \wedge g);$ and the sequence $id(H_1) + +id(H_2) + +id[r] \not\in P;$
and $P' = P \cup id(H_1) + +id(H_2) + +[r]$
\end{tabular}}
\end{center}

\vspace{4mm}
{\bf Example 3.1}
\label{leqwt}
The following is a (terminating) derivation under $\omega_t$ for the query
$leq(A,B), leq(B,C), leq(C,A)$ executed on the leq program in Example
\ref{chr}. For brevity, $P$ have been removed from each tuple.

\vspace{4mm}
\scalebox{0.65} {
\begin{tabular}{p{4.6cm} p{1cm} l p{0.1cm}}
&&$\langle\{leq(A,B),leq(B,C),leq(C,A)\},\emptyset,\emptyset\rangle_1$&(1)
\\

&&&\\

&$\mapsto_{introduce}$&$\langle\{leq(B,C),
leq(C,A)\},\{leq(A,B)\#1\},\emptyset\rangle_2$&(2) \\

&&&\\

&$\mapsto_{introduce}$&$\langle\{leq(C,A)\},\{leq(A,B)\#1,
leq(B,C)\#2\},\emptyset\rangle_3$&(3)\\

&&&\\

(transitivity r4 $X=A \wedge Y=B \wedge
Z=C$)&$\mapsto_{apply}$&$\langle\{leq(C,A),leq(A,C)\},\{leq(A,B)\#1,leq(B,C)\#2\},\emptyset\rangle_3$&(4)
\\

&&&\\

&$\mapsto_{introduce}$&$\langle\{leq(C,A)\},\{leq(A,B)\#1,leq(B,C)\#2,
leq(A,C)\#3\},\emptyset\rangle_4$&(5)
\\

&&&\\

&$\mapsto_{introduce}$&
$\langle\emptyset,\{leq(A,B)\#1,leq(B,C)\#2,leq(A,C)\#3,
leq(C,A)\#4\},\emptyset\rangle_5$&(6)
\\

&&&\\

(antisymmetry r2 $X=C \wedge Y=A $)&$\mapsto_{apply}$&$\langle\emptyset,
\{leq(A,B)\#1, leq(B,C)\#2\},\{A=C\}\rangle_5$&(7)
\\

&&&\\

(antisymmetry r2 $X=C \wedge Y=A $)&$\mapsto_{apply}$&$\langle\emptyset,
\emptyset,\{A=C, C=B\}\rangle_5$&(8)
\\

\end{tabular}
} 

\vspace{3mm}
\normalsize{No more transition rules are possible, so this is the final state.}

\subsubsection{Theoretical Operational Semantics $\omega_t$ of CHR in SFC}
\label{sec:tosomtChr}

The following is the description of the theoretical operational semantics $\omega_t$ in terms of the sorts, relations, functions and axioms of the simple fluent calculus. 

\vspace{2mm}
\begin{enumerate}[(a)]
  \item Domain Sorts
  \begin{enumerate}[-]
	  \item $NATURAL$, natural numbers;
	  \item $RULE$, the sort of CHR rules and $RULE\_ID$ the sort of the rule
	  identifiers;
	  \item $CONSTRAINT$, the sort of constraints, with the following subsorts:
	  $BIC$ (the built-in constraints), with the subsort $EQ$ (constraints in the
	  form $x = y$), and $UDC$ (the user-defined constraints), with the following
	  subsort: $IDENTIFIED$ (constraints in the form $c \# i$). In short: 
          
$EQ < BIC  < CONSTRAINT$ and 

$IDENTIFIED < UDC < CONSTRAINT$;
	  \item $PROPHISTORY = Seq(NATURAL) \times RULE$, the elements of the
	  Propagation History, tuples of a sequence of natural numbers and a rule;
	  \subitem For each defined sort $X$, three new sorts: $Seq(X)$, $Set(X)$ and
	  $Bag(X)$ containing the sequences, the sets and the bags of elements of $X$.
	  We use [] for the empty sequence and \{\} for the empty set and the empty bag.  
	  \item $CHR ACTION < ACTION$, the subsort of $ACTION$ containing only the
	  actions in the CHR semantics.    
  \end{enumerate}
  \item Predicates
  \begin{enumerate}[-]
	  \item $Query : Bag(CONSTRAINT)$, $Query(q)$ holds iff $q$ is the initial
	  query;
	  \item $Consistent : STATE$, holds iff the $BICS$ of the state is consistent
	  (i.e., if it does not entail false).
	  \item $Match(h_k, h_R, u_1, u_2, e, z)$ holds iff (i) $u_1$ and
	  $u_2$ are in the $UDCS$ of $z$ and (ii) the set of matching equations e is
	  such that $chr(u_1) = e(h_k)$ and $chr(u_2) = e(h_R)$;
	  \item $Entails : Set(BIC) \times Set(EQ) \times Bag(BIC)$, $Entails(b, e, g)$
	  holds if $CT \models b \rightarrow \exists (e \wedge g)$.    
  \end{enumerate}  
  \item Functions
  \begin{enumerate}[-]
	  \item $\# : UDC \times NATURAL \mapsto IDENTIFIED$, defines the syntactic
	  sugar for defining identified constraints in the form $c\#i$;   
	  \item $makeRule : $

$RULE ID \times Bag(UDC) \times Bag(UDC) \times Bag(BIC) \times Bag(UDC) \mapsto RULE$, 

makes a rule from its components. We
	  define the syntactic sugar for rules as $r_{id} @ h_k \backslash h_R
	  \leftrightarrow g|b = makeRule(r_{id}, h_k, h_R, g, b)$;
	  \item $Bics : STATE \mapsto Set(BIC)$, where $Bics(z) =
	  \left\{c|Holds(InBics(c), z)\right\}$;
	  \item $id : Set(UDC) \mapsto Set(NATURAL)$, where $id(H) = {i|c\#i \in H}$
	  \item The usual set, sequence and bag operations: $\in$ for element, $\cup$
	  for set union, $\uplus$ for bag union, $++$ for sequence concatenation, $|$
	  for sequence head and tail (Ex: $ [ head|tail] $) and $ \backslash $ for set 
	  subtraction.
  \end{enumerate}    
  \item Fluents
  \begin{enumerate}[-]
	  \item $Goal : Bag(UDC) \mapsto FLUENT$, $Goal(q)$ holds iff $q$ is the
	  current goal;
	  \item $Udcs : Bag(IDENTIFIED) \mapsto FLUENT$, $Udcs(u)$ holds iff $u$ is the
	  current UDCS;
	  \item $InBics : BIC \mapsto FLUENT$, $InBics(c)$ holds iff $c$ is in the
	  current BICS;
	  \item $InPropHistory : PROPHISTORY \mapsto FLUENT$, $InPropHistory(p)$ holds
	  iff $p$ is in the current Propagation History;
	  \item $NextId : NATURAL \mapsto FLUENT$, $NextId(n)$ holds iff $n$ is the
	  next natural number to be used to identify a identified constraint.    
  \end{enumerate} 
  
\item Actions 
  \begin{enumerate}[-]
	  \item $Solve : BIC \mapsto CHR\_ACTION$, $Do(Solve(c), s)$ executes the
	  $Solve$ transition with the built-in constraint $c$;
	  \item $Introduce : UDC \mapsto CHR\_ACTION$, $Do(Introduce(c), s)$ executes
	  the Introduce transition with the user-defined constraint $c$;
	  \item $Apply : RULE \times Bag(UDC) \times Bag(UDC) \mapsto CHR\_ACTION$,
	  
$Do(Apply(r, u_1, u_2), s)$ executes the Apply transition matching the
	  constraints $u_1$ and $u2$ in the $UDCS$ with the kept and removed heads of
	  $r$.
  \end{enumerate}  
   
  \item Axioms
  \begin{enumerate}[-]
	  \item $Query(q) \rightarrow State(S0) = Goal(q) \circ Udcs(\left\{\right\})
	  \circ NextId(1)$, 
	  
	  The Initial State Axiom states that in the initial state,
	  the goal contains the constraints in the query, the user defined constraint
	  store is empty and the next ID for identified constraints is 1;

\vspace{4mm}
\texttt{Solve}

	  \item $Poss(Solve(c), z) \equiv (\exists q)(Holds(Goal(q), z) \wedge c \in
	  q)$  
	 
	 The Solve Precondition Axiom states that the only precondition for the
	 $Solve$ action on the built-in constraint $c$ is that this constraint should
	 be in the goal.
	 \item $Poss(Solve(c), s) \wedge Holds(Goal(q \uplus \left\{c\right\}),
	 State(s)) \supset  $

$\ \ \ \ State(Do(Solve(c), s)) = State(s) \circ Goal(q) \circ 
	 InBics(c) -  $

$\ \ \ \ Goal(q \uplus \left\{c\right\})$ 
	
	The Solve State Update Axiom states that the result of the $Solve$ action over
	the constraint $c$ is that this constraint is removed from goal and added to
	$InBics$ list in current state;

\vspace{4mm}
\texttt{Introduce}

	\item $Poss(Introduce(c), z) \equiv (\exists q)(Holds(Goal(q), z) \wedge c \in
	q)$
	\item $Poss(Introduce(c), s) \wedge Holds(Goal(q \uplus {c}), State(s))
\wedge $

$\ \ \ \ Holds(Udcs(u), State(s)) \wedge Holds(NextId(n), State(s))
\supset $

$\ \ \ \ State(Do(Introduce(c), s)) = State(s) \circ Goal(q) \circ Udcs(u \uplus {c\#n})
\circ NextId(n + 1) - $

$\ \ \ \ Goal(q \uplus \left\{c\right\}) - Udcs(u) - NextId(n)$ 

\vspace{4mm}
\texttt{Apply}

     \item $Poss(Apply(r @ hk \backslash hR \leftrightarrow g|d, u_1, u_2), z)
    \equiv $

$\ \ \ \ (\exists e)(\exists b)(Match(h_k, h_R, u_1, u_2, e, z) \wedge$

$\ \ \ \ \lnot Holds(InPropHistory(id(u_1), id(u_2), r), z) \wedge Bics(b, z) \wedge  Entails(b, e, g))$
    \item $Poss(Apply(r @ hk \backslash hR \leftrightarrow g|d, u_1, u_2),
    State(s)) \wedge $

$\ \ \ \ Holds(Udcs(u_1 \uplus u_2 \uplus u), State(s))
\wedge Holds(Goal(q), State(s)) \wedge $

$\ \ \ \ Match(h_k, h_R, u_1, u_2, e, z) \supset $

$\ \ \ \ State(Do(Apply(r @ hk \backslash hR \leftrightarrow g|d, u_1, u_2),
s)) = State(s) \circ Goal(d \uplus q) \circ $

$\ \ \ \ Udcs(u1 \uplus u) \circ InBics(e)
\circ InBics(g) \circ InInPropHistory(id(u_1), id(u_2), r)
- $

$\ \ \ \ Goal(q) - Udcs(u1 \uplus u2 \uplus u)$
  \end{enumerate}   
\end{enumerate} 

\subsubsection{Observational Semantics of CHR based on $\omega_t$}
\label{sec:obsemChr}

The following is the description of the observational semantics of CHR using the simple fluent calculus with modified axioms of the Section~\ref{sec:obssemSfc}. Sorts, Predicates, Functions and Fluents are the same as in the previous section; there is on additional item for the attributes.

\vspace{2mm}
The actions are now constants and we make explicit 4 actions: 

$\ \ \ \ \ \ \ \ \ \ Init, Solve, Introduce, Fail$. 
\begin{enumerate} [ ]

 \item (e) Actions 
  \begin{enumerate}[-]
          \item $Init : \mapsto CHR\_ACTION$, $Do(Init, [goal(q)|a], s)$ executes the
	  top-level initial transition (starting the resolution) with some query $q$ in the current state ($a$ stands for other attributes list in the associated trace event);
	  \item $Solve : \mapsto CHR\_ACTION$, $Do(Solve,[bic(c)|a], s)$ executes the
	  $Solve$ transition with the built-in constraint $c$;
	  \item $Introduce : \mapsto CHR\_ACTION$, $Do(Introduce, [udc(c)|a], s)$ executes
	  the Introduce transition with the user-defined constraint $c$;
	  \item $Apply : \mapsto CHR\_ACTION$,
	  
           $Do(Apply, [rule(r)|t], s)$ executes the Apply transition with rule $r$ matching the
	  constraints in the $UDCS$ with the kept and removed heads;
	  \item $Fail : \mapsto CHR\_ACTION$, $Do(Fail, [goal(q)|a], s)$ if no $Apply$ is possible.
  \end{enumerate}

\vspace{2mm}
There are also 5 attributes in the actual trace: $goal, udc, bic, hind, rule$.

  \item (f) Attributes
  \begin{enumerate}[-]
	  \item $goal : CONSTRAINTS \mapsto ATTRIBUTE$, is the set of constraints in the current Goal;
	  \item $udc : CONSTRAINTS \mapsto ATTRIBUTE$, is the set of constraints in the current User Defined Constraints Store;
	  \item $bic : CONSTRAINTS \mapsto ATTRIBUTE$, is the set of constraints in the current Built-In Constraints Store ;
	  \item $hind : \mapsto INTEGER$, is the new propagation history index (incremented by $Introduce$);
	  \item $rule: RULE \mapsto ATTRIBUTE$, is the rule applied to reach this state.
  \end{enumerate} 

\vspace{4mm}
\texttt{Apply}

We just comment the adaptation of one rule, the full description is in Appendix~\ref{app:obssemCHRor}.

\item (g) {Axioms of the Observational Semantics}

   \begin{enumerate}[-]
    \item $Poss(Apply, [r, h_k, h_R, g, u_1, u_2], z)  \equiv $

$\ \ \ \ (\exists e)(\exists b)(Match(h_k, h_R, u_1, u_2, e, z) \wedge$
    
$\ \ \ \ \lnot Holds(InPropHistory(id(u_1), id(u_2), r), z) \wedge Bics(b, z) \wedge $

$\ \ \ \  Entails(b, e, g))$

    \item $Poss(Apply, [r, h_k, h_R, g, u_1, u_2], State(s)) \wedge $

$\ \ \ \ Holds(Udcs(u_1 \uplus u_2 \uplus u), State(s))
\wedge Holds(Goal(q), State(s)) \wedge $

$\ \ \ \ Match(h_k, h_R, u_1, u_2, e, z) \supset $

$\ \ \ \ State(Do(Apply, $

\ \ \ \ [apply, $rule(r @ h_k \backslash h_R \leftrightarrow g|d, u_1, u_2), goal(d \uplus q), udc(u1 \uplus u), bic(g)$]$, s)) =$

$\ \ \ \ State(s) \circ Goal(d \uplus q) \circ Udcs(u1 \uplus u) \circ InBics(e) \circ InBics(g) \circ $

$\ \ \ \ \ \  InPropHistory(id(u_1), id(u_2), r) $

$\ \ \ \ \ \  - Goal(q) - Udcs(u1 \uplus u2 \uplus u)$
 \end{enumerate}  
\end{enumerate}

\vspace{4mm}
The observational semantics of CHR$^\vee$ can be formalized similarly with 11 actions (hence 11 different kinds of trace events), 

$initState, solve, activate, reactivate, drop, simplify, propagate, derive,$

$clean, split, fail$, 

\noindent
using the refined operational semantics $\omega_r^{\vee}$, according to \cite{duck2004refined,silva2009ufpe}.

\subsection{Towards Full Generic Trace of CHR$^\vee$}
\label{sec:compoCHRsem}

As suggested in the Section~\ref{sec:compos}, a full trace will be progressively obtained by composing several layers of traces and several potential applications in such a way that as many as possible of potential uses can be satisfied by such a trace. The Figure~\ref{fig:chromepipiline} suggests 4 levels of refinements corresponding to 4 layers of implementations, i.e. from bottom to top: environment of execution (Windows/Linux/Mac...), implementation language (most of CHR are implemented in Prolog), CHR, and application written in CHR. There may be other lower levels, like WAM abstract machine implemented in Java for the Prolog level, etc... Even if each layer has its own level of abstraction and most of the CHR users don't care about lower software layers, it may be interesting to keep some trace of them in the ``full'' trace. 

If we consider the point of view of debugging some application written in CHR, here are some information which could be usefully found in a trace used by a debugging tool.

\begin{itemize}
\item Execution environment:  activation of system commands during interactions
\item Implementation languages (there may be several layers): specific local error messages, activated layer, ...
\item CHR: name of used rules
\end{itemize}

We mean here that, at some point, it may be useful to find in the trace of the application some information regarding different layers of implementation in order a debugging tool to be able to ``understand'' some bugs.

\vspace{2mm}
Let us consider an example of application.
In the Annex~\ref{app:robots}  we give the observational semantics in SFC of a simple application of \cite{Thielscher98}. This OS specifies possible traces of actions performed by robots (here there is only one). We may assume that this small world is programmed in CHR and therefore the trace of the whole system is a kind of combination of both traces: the one of the robot and the trace corresponding to the CHR program execution. The resulting full trace corresponds to the trace composition described in the Section~~\ref{sec:compos} (comprehensively treated in \cite{TMTmanuscript}).

If one wants just to follow what the robots are doing, then the sub-trace consisting of the trace events regarding the robot's actions is sufficiently relevant. But, at least at the stage of debugging, some dysfunction observed in the robot's trace (for example crossing a closed door) can be understood only by looking at a more complete trace which includes events related to the CHR layer behaviour.

\vspace{3mm}
Finally there is one more level, which corresponds to the specificity and versatility of CHR: the many extensions and applications which are embedded in CHR with CHR as implementation language, quoted as the ``CHR world'' in the introduction. Here are some of them \cite{fruhwirth2003essentials}: Boolean algebra for circuit analysis, resolution of linear polinomial equations - CLP($\cal{R}$)\footnote{CLP stands for Constraint Logic Programming.}- with application in finances and non linear equations, finite domain solvers - CLP(FD) - with applications in puzzles, scheduling and optimisation, but many others as quoted at the beginning of the report.

\vspace{1mm}
A full CHR$^\vee$ trace should probably include traces related to several extensions like CLR($\cal{X}$) where $\cal{X}$ stands for some constraint domain, and CHR$^{V;naf}$ for example. This is possible, but there is still a need to specify an observational semantics for several of these extensions.


\newpage
\section{Towards CHROME-REF}
\label{chromeref}

This section details the architecture of CHROME-REF, the extensible implementation of a generic tracer for Constraint Solving and Rule-Based Reasoning.
Each component of the CHROME-REF is described
in terms of UML2.1 according the KobrA2 methodology \cite{atkinson2000component,robin2009kobra2}. 
We first give a brief overview of CHROME.

\subsection{CHROME}
\label{sec:chrome}

CHROME stands for \textbf{Constraint Handling Rule Online Model-driven Engine},
is a model-driven, component-based, scalable, online, Java-hosted CHR$^\vee$
engine to lay at the bottom of the framework as the most widely reused automated reasoning
component. The idea of CHROME is also to demonstrate how a standard set of
languages and processes prescribed by MDA can be used to design concrete
artefacts, such as: a versatile inference engine for CHR$^\vee$ and its
compiler component that generates from a CHR$^\vee$ base the source code of Java
classes.

\subsubsection{Goals and Design Principles}
\label{sec:chromegoals}

The main goal of CHROME if to take CHR engines a step beyond, by
designing a new CHR$^\vee$ engine and a corresponding compiler using a
component-based model-driven approach. CHROME is a CHR$^\vee$ engine with an
efficient and complete search algorithm (e.g. the conflict-directed
backjumping algorithm), the first versatile rule-based engine, integrating
production rules, rewrite rules, its built-in belief revision mechanism
(reused for handling disjunctions) and CLP rules to run on top of a mainstream
Object Oriented (OO) platform (Java). Because it is a rule-based engine following a
component-based model-driven approach, it allows easy port to other
OO platforms such as Python, JSP, C++ and others. Finally the compiler is the
first that uses a model transformation pipeline to transform from a source
relational-declarative language into a OO imperative paradigm language.

\begin{figure} 
\centering
\scalebox{0.8}{\includegraphics{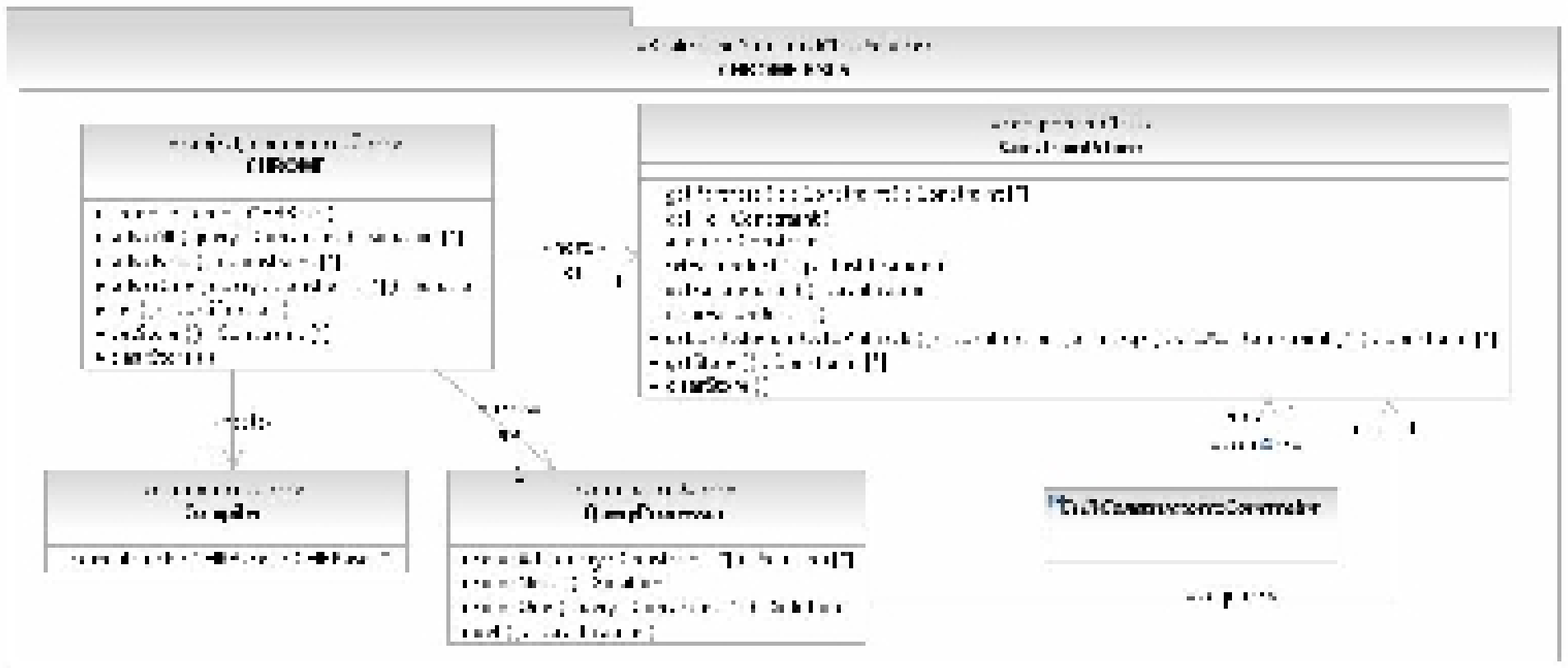}}
\caption[CHROME]
{CHROME}
\label{fig:chromerscs}
\end{figure}

The CHROME architecture is divided into two main sub-components (see the 
Figure~\ref{fig:chromerscs}):

\begin{enumerate}[i)]
  \item The ATL-pipeline compiler (CHR Compiler component) that takes as input a
  relational declarative CHR$^\vee$ base and produces an efficient constraint
  handling imperative object-oriented component assembly.
  \item The CHROME run-time engine (shown in the Figure~\ref{fig:chromerscs} as the QueryProcessor
  component) that provides the services and data structures necessary to
  execute a CHR$^\vee$ base given a particular query (collection of constraints).   
\end{enumerate}

\begin{figure} 
\centering
\scalebox{0.7}{\includegraphics{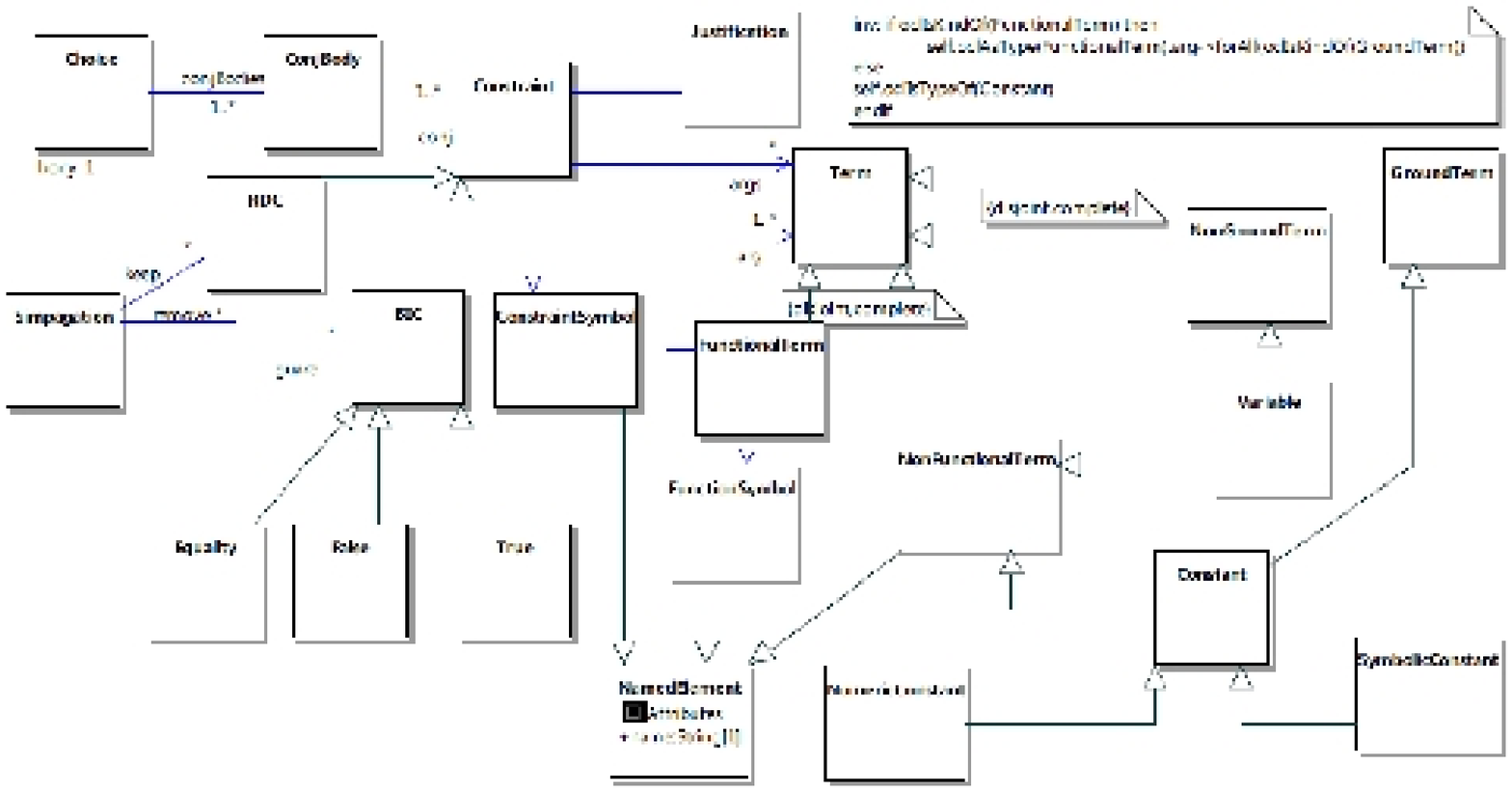}}
\caption[Meta-model of CHR$^\vee$ and CHR data structures]
{Meta-model of CHR$^\vee$ and CHR data structures.}
\label{fig:chrmetamodel2}
\end{figure}

\vspace{3mm}
The Figure~\ref{fig:chrmetamodel2} shows the complete MOF metamodel of CHR$^\vee$. At this
abstract syntax level all CHR$^\vee$ rules are generalized as simpagation rules. The meta-associations
keep and del from the CHR$^\vee$ meta-class to the Constraint meta-class
respectively represent the propagated and simplified heads of the rule. The
heads must be instances of RDCs (Rule Defined Constraints). A guard of a rule
must be a collection of BICs (BuiltIn Constraints). Both RDCs and BICs are
specializations of Constraint meta-class.

\vspace{3mm}
The body of a CHR$^\vee$ rule is a collection of alternative conjunctions.
Conjunctions are composed by both RDC or BIC, e.g. a collection of instances
of meta-class Constraint. The original CHR$^\vee$ base has a special RDC (OR) to
indicate disjunctions in the body. The collection of all rules of a CHR
program is a CHR Base.

\vspace{1mm}
Each constraint is composed by a constraint symbol and a collection of zero or
more arguments, e.g. a collection of terms (meta-class Term). A Term further
specializes into: functional terms, non-functional terms, ground terms and
non-ground terms. A Constant is both a non-functional term and a ground term
and a Variable is a non-ground term and a non-functional term. Finally a
functional term is further composed by a Function Symbol and a collection of
zero or more arguments, which are in turn recursively defined as instances of
meta-class Term. The constraint domain meta-class aggregates all term symbols
allowed.

The metamodel displays also the internal structures of the engine, namely: the
constraint store and the constraint queue. The first stores the constraints
added by firing rules, the second is a processing queue that tracks which is
the next constraint to be processed.

\subsubsection{Strengths and Limitations}
\label{sec:chromestrengthsandlimitations}

CHROME is the first Java CHR$^\vee$ engine: none of the related CHR Java engines allow disjunctive rules.
Compilation in Java makes it easier to reuse and deploy full CHR$^\vee$ bases in applications in need of automated reasoning services.
It is one of the largest case study to date to integrate MDE
technology with model transformations (4358 ATL lines) and components. 
It however suffers some limitations, as it provides only three built-in constraints: the syntactical equality, true and false, and it has no visual tracing IDE. This makes practical applications still too cumbersome to
implement, being tracing a fundamental part of large automated reasoning development.

\subsection{CHROME-REF Components}
\label{sec:traceschema}

\begin{figure} 
\centering
\scalebox{0.6}{\includegraphics{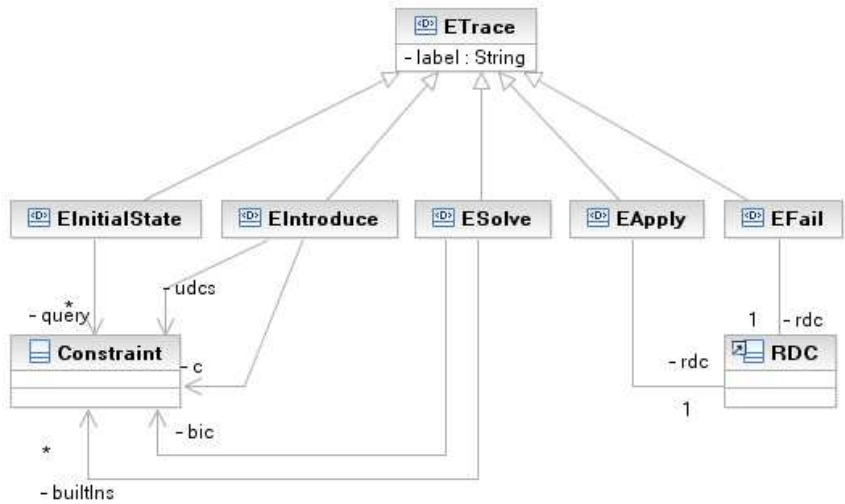}}
\caption[OO Representation of the OS of the CHR$^\vee$ Tracer]
{OO Representation of the OS of the CHR$^\vee$ Tracer}
\label{fig:traceschema}
\end{figure}

The Figure~\ref{fig:traceschema} shows a object-oriented representation of
the observational semantics of CHR$^\vee$ as described in the Section~\ref{sec:obsemChrSfc}. It contains five sort of extracted trace events ($ETrace$): an initial
state ($EInitialState$), user-defined contraint store introduced
($EIntroduce$), built-in introduced ($ESolve$), rule applied ($EApply$) and
rule failed ($EFail$).

\begin{figure} 
\centering
\scalebox{0.4}{\includegraphics{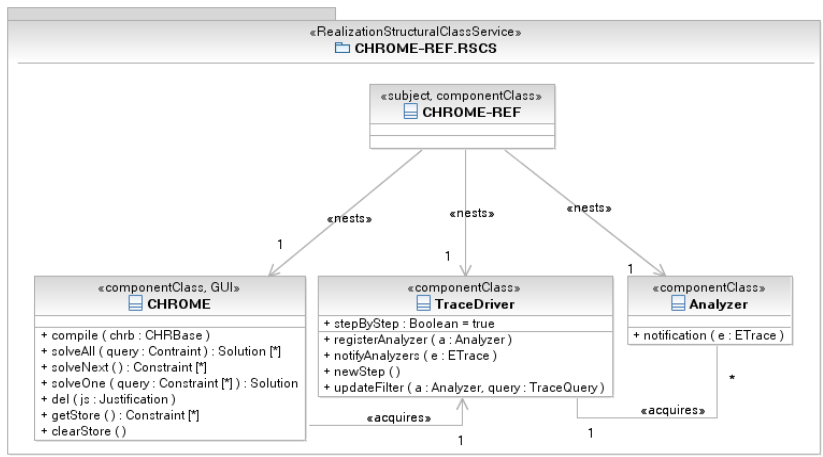}}
\caption[Meta-model of CHR$^\vee$ and CHR data structures]
{CHROME-REF three main Components}
\label{fig:chromerefrscs}
\end{figure}

The top-level CHROME-REF component encapsulates all sub-components that compose
the CHROME environment. The Figure \ref{fig:chromerefrscs} shows the main
component and its three sub-components as defined in what follows. 
They provide methods to
compile a rule base, to solve a query (displaying one or more solutions for
such query), to adapt a solution when a given set of justified constraints is
deleted and to clear the constraint store for processing a new query. 

The Driver component is a intermediator between CHROME and Analyzer, its function is
to manager the communication of trace event sent by the engine and filter the
requested information to the analyzer. Finally, the Analyzer component, in our
case, is a debugging tool for CHR.

The next sub-sections describe each component.

\subsubsection{Extraction}



\begin{figure} 
\centering
\scalebox{0.3}{\includegraphics{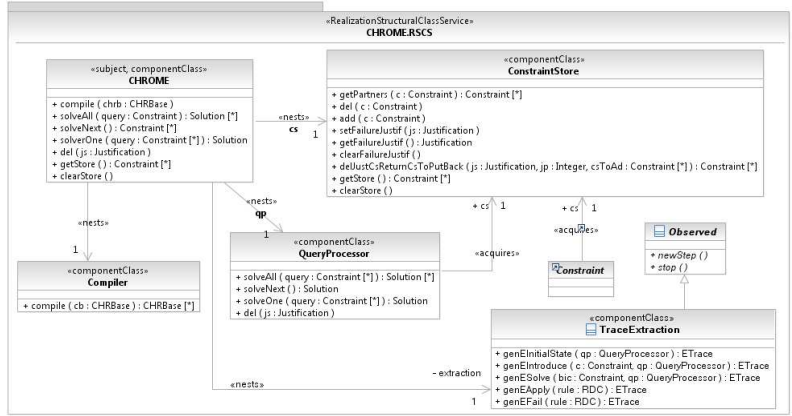}}
\caption[Insertion of the Trace Extraction Component into CHROME]
{Insertion of the Trace Extraction Component into CHROME}
\label{fig:chrometracerscs}
\end{figure}

The Figure \ref{fig:chrometracerscs} shows the CHROME
component with a new component called {\tt TraceExtraction}, which implements the generic CHR trace as described in the Section \ref{sec:obsemChr}.

We have added two improvements to the CHROME to integrate with our proposal: a
new componente called Trace Extraction that receives as input some
parameters ($Constraint, QueryProcessor and RDC$) and produces the trace events
($ETrace$); and, we included new rules into the CHROME compiler to send the
previous parameters to the Trace Extraction during the execution of a CHR
program.

The following OCL rules explain how a trace event will be produced from received
parameters:

\lstset{
	backgroundcolor==\color{darkgray},
	tabsize=4,
	rulecolor=,
	language=OCL,
        basicstyle=\footnotesize,  
        upquote=true,
        aboveskip={1.5\baselineskip},
        columns=fixed,
        showstringspaces=false,
        extendedchars=false,
        breaklines=true,
        prebreak = \raisebox{0ex}[0ex][0ex]{\ensuremath{\hookleftarrow}},
        showtabs=false,
	numberstyle=\tiny,
	numbersep=5pt,
    showspaces=false,	
    showstringspaces=false,
    identifierstyle=\ttfamily,
    keywordstyle=\color{black}\bfseries\emph,
    commentstyle=\color[rgb]{0.133,0.545,0.133},
    stringstyle=\color[rgb]{0.627,0.126,0.941},
}
\begin{lstlisting}
context TraceExtraction::genEInitialState(qp:QueryProcessor):Etrace
post:
	let eInitialState:EInitialState = oclIsNew()
	in 
		eInitialState.query = qp.goal and
		result = eInitialState

context TraceExtraction::genEIntroduce(c:Constraint, qp:QueryProcessor):Etrace
post: 
	let eIntroduce:EIntroduce = oclIsNew()
	in 
		eIntroduce.c = c and
		eIntroduce.udcs = qp.cs.getStore() and
		result = eIntroduce

context TraceExtraction::genESolve(bic:Constraint, qp:QueryProcessor):Etrace
post:
	let eSolve:ESolve = oclIsNew()
	in 
		eSolve.bic = bic and
		eSolve.builtIns = qp.allVars and
		result = eSolve

context TraceExtraction::genEApply(rule:RDC):Etrace
post: 
	let eApply:EApply = oclIsNew()
	in 
		eApply.rule = rule and
		result = eApply

context TraceExtraction::genEFail(rule:RDC):Etrace
post:
	let eFail:EFail = oclIsNew()
	in 
		eFail.rule = rule and
		result = eFatil
\end{lstlisting}

\subsubsection{Driver}

\begin{figure} 
\centering
\scalebox{0.6}{\includegraphics{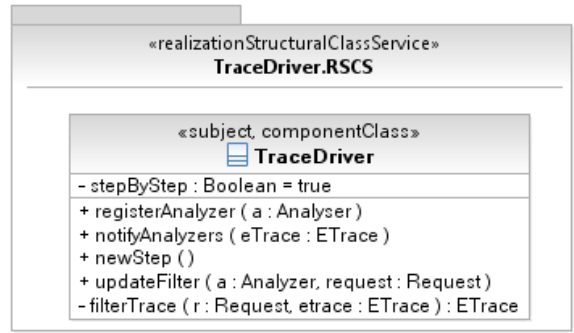}}
\caption[Trace Driver Meta-model]
{Trace Driver Meta-model}
\label{fig:tracedriverrscs}
\end{figure}

The Trace Driver component is an intermediator between a process and an
analyzer. It component has the following functions (Figure~\ref{fig:tracedriverrscs}):

\begin{itemize}
  \item to decide whether the underlying process (in our case CHROME) will send
  trace events step by step or all at once. This information is represented by
  means of the flag $stepByStep$;
  \item to register an analyzer that will watch the trace events from the
  underlying process;
  \item to notify all connected analyzers soon after a trace event to be produced;
  \item to ask for a new trace event (only if the flag $stepByStep$ is true);
  and,
  \item to filter a trace event by means of a trace query sent from a analyzer.
\end{itemize}

The following OCL rules describe the post-condition of each method:

\lstset{
	backgroundcolor==\color{darkgray},
	tabsize=4,
	rulecolor=,
	language=OCL,
        basicstyle=\footnotesize,  
        upquote=true,
        aboveskip={1.5\baselineskip},
        columns=fixed,
        showstringspaces=false,
        extendedchars=true,
        breaklines=true,
        prebreak = \raisebox{0ex}[0ex][0ex]{\ensuremath{\hookleftarrow}},
        showtabs=false,
	numberstyle=\tiny,
	numbersep=5pt,
    showspaces=false,	
    showstringspaces=false,
    identifierstyle=\ttfamily,
    keywordstyle=\color{black}\bfseries\emph,
    commentstyle=\color[rgb]{0.133,0.545,0.133},
    stringstyle=\color[rgb]{0.627,0.126,0.941},
}
\begin{lstlisting}
context TraceDriver::registerAnalyzer(a:Analyzer)
post: analyzer->includes(d)

context TraceDriver::notifyDriver(eTrace:ETrace)
post: analyzers->forAll(a | a^notification(filterTrace(a.request, eTrace)))

context TraceDriver::newStep()
post: observed.newStep()

context TraceDriver::updateFilter(a:Analyzer, request:Request)
post: a.request = request
\end{lstlisting}

\begin{figure} 
\centering
\scalebox{0.45}{\includegraphics{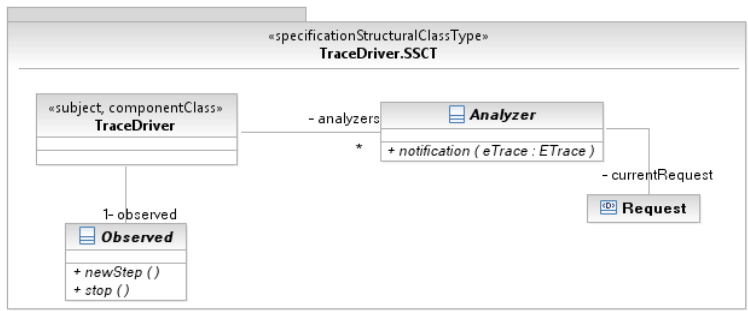}}
\caption[Meta-model of the Analyzer and Communication with the Driver]
{Meta-model of the Analyzer and Communication with the Driver}
\label{fig:tracedriverssct}
\end{figure}

As said before, the Trace Driver component is a intermediator that receives
trace events from a process and sends it to the analyzer. The Figure
\ref{fig:tracedriverssct} shows the relationships between these components.  

A trace driver is connected to an observed process which sends the trace events,
and it has a list of analyzers to which to  the trace events. Each analyzer is
associated to a query, which says which information the
driver will send to the analyzer.

\subsubsection{Analyzer}
\label{sec:analyser}

The trace analyzer specifies to the driver which events are
needed by means of queries. The requests that an analyzer can send to the
tracer driver are of three kind. Firstly, the analyzer can ask for additional
data about the current event. Secondly, the analyzer can modify the query to
be checked by the driver. Thirdly, the analyzer can notify the driver to
pause, continue or end the process execution.  

\begin{figure} 
\centering
\scalebox{0.3}{\includegraphics{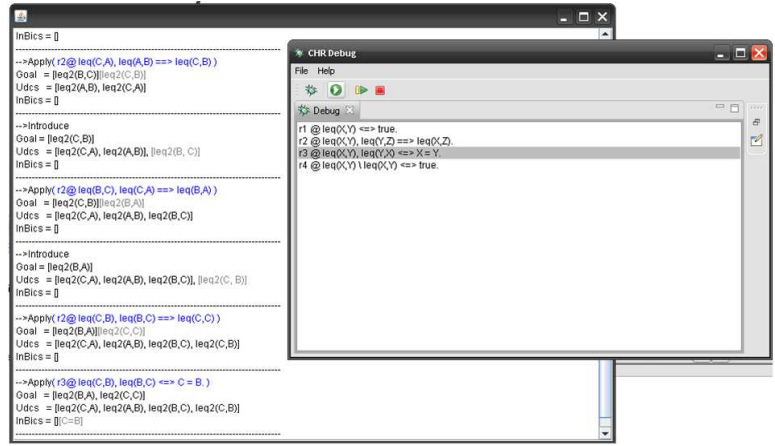}}
\caption[Meta-model of CHR$^\vee$ and CHR data structures]
{A Simple GUI to Analyze a Program Execution Looking at its Trace}
\label{fig:leqdebug}
\end{figure}

In this project, in order to exemplify the whole proposed framework, we have
created two simple views to show a pretty-printing of a CHR execution. The
Figure \ref{fig:leqdebug} presents these two kind of analyzers: the $Trace \ View$
shows the evolution of the CHR parameters (Goal, Constraint Store, Built-ins,
etc), defined in the generic trace; and the
$Program\ View$ is just to focus in a specific rule when this rule is triggered.

\vspace{2mm}
We illustrate these views with an execution of the LEQ example.

\lstset{
	backgroundcolor==\color{darkgray},
	tabsize=4,
	rulecolor=,
	language=Prolog,
        basicstyle=\footnotesize,  
        upquote=true,
        aboveskip={1.5\baselineskip},
        columns=fixed,
        showstringspaces=false,
        extendedchars=true,
        breaklines=true,
        prebreak = \raisebox{0ex}[0ex][0ex]{\ensuremath{\hookleftarrow}},
        showtabs=false,
	numberstyle=\tiny,
	numbersep=5pt,
    showspaces=false,	
    showstringspaces=false,
    identifierstyle=\ttfamily,
    keywordstyle=\color{black}\bfseries\emph,
    commentstyle=\color[rgb]{0.133,0.545,0.133},
    stringstyle=\color[rgb]{0.627,0.126,0.941},
}
\begin{lstlisting}
reflexivity  @ leq(X,Y) <=> X=Y | true.
antisymetry  @ leq(X,Y) , leq(Y,X) <=> X=Y.
idempotence  @ leq(X,Y) \ leq(X,Y) <=> true.
transitivity @ leq(X,Y) , leq(Y,Z) <=> leq(X,Z).
\end{lstlisting}

with the query:
\lstset{
	backgroundcolor==\color{darkgray},
	tabsize=4,
	rulecolor=,
	language=Prolog,
        basicstyle=\footnotesize,  
        upquote=true,
        aboveskip={1.5\baselineskip},
        columns=fixed,
        showstringspaces=false,
        extendedchars=true,
        breaklines=true,
        prebreak = \raisebox{0ex}[0ex][0ex]{\ensuremath{\hookleftarrow}},
        showtabs=false,
	numberstyle=\tiny,
	numbersep=5pt,
    showspaces=false,	
    showstringspaces=false,
    identifierstyle=\ttfamily,
    keywordstyle=\color{black}\bfseries\emph,
    commentstyle=\color[rgb]{0.133,0.545,0.133},
    stringstyle=\color[rgb]{0.627,0.126,0.941},
}
\begin{lstlisting}
leq(A,B), leq(B,C), leq(C,A).
\end{lstlisting}

Finally, we get the XML instance produced by CHROME for this example (see complete trace with all attributes in the Appendix~\ref{app:leqexample}).

\lstset{
	backgroundcolor==\color{darkgray},
	tabsize=4,
	rulecolor=,
	language=Prolog,
        basicstyle=\footnotesize,  
        upquote=true,
        aboveskip={1.5\baselineskip},
        columns=fixed,
        showstringspaces=false,
        extendedchars=true,
        breaklines=true,
        prebreak = \raisebox{0ex}[0ex][0ex]{\ensuremath{\hookleftarrow}},
        showtabs=false,
	numberstyle=\tiny,
	numbersep=5pt,
    showspaces=false,	
    showstringspaces=false,
    identifierstyle=\ttfamily,
    keywordstyle=\color{black}\bfseries\emph,
    commentstyle=\color[rgb]{0.133,0.545,0.133},
    stringstyle=\color[rgb]{0.627,0.126,0.941},
}
\begin{lstlisting}
<?xml version="1.0" encoding="UTF-8"?>
<chrv
	xmlns="http://orcas.org.br/chrv"
	xmlns:xsi="http://www.w3.org/2001/XMLSchema-instance" 
	xsi:schemaLocation=
		"http://orcas.org.br/chrv chrv.xsd">	 
	<event chrono="1">
		<initialState>
			<goal>leq(A,B), leq(B,C), leq(C,A)</goal>
		</initialState>
	</event>
	<event chrono="2">
		<introduce>
			<udc>leq(A,B)</udc>
			<goal>leq(B,C), leq(C,A)</goal>
		</introduce>
	</event>
	<event chrono="3">
		<introduce>
			<udc>leq(A,B), leq(B,C)</udc>
			<goal>leq(C,A)</goal>
	...															
</chrv>
\end{lstlisting}

At this stage of the implementation, such views are principally helpful to help to develop the generic trace.


\newpage
\section{Conclusion}
\label{conclusion}

In this report we have presented an ongoing work as a roadmap towards CHROME-REF, and we have defined the methods and tools to reach this goal. They consists of a formal specification (called observational semantics - OS) of a generic tracer for CHR$^\vee$ using an adaptation of the simple fluent calculus (SFC) which we have presented, and its implementation, over and inside CHROME, using the KobrA2 method, and resulting in a PIM of CHROME-REF. 

We have tested the approach with a small trace pretty printing GUI. We also indicated how to work on extensions of the very first trace we presented (using the simple $\omega^t$ semantics of CHR), before including other actions and attributes inspired by more refined semantics of CHR$^\vee$ and various CHR domains extensions.

\vspace{2mm}
Several other issues however remain to be explored.

\vspace{1mm}
First the use of the SFC to describe the observational semantics of tracers. As quoted in the report, we may expect several advantages of its use: facilitation of trace extension specification, or refinement, by merging observational semantics of several CHR extensions or sublayers, and facilitation of verification of formal properties of the trace. We did not reach the point to be in condition to simulate production of traces trying to execute this OS using Flux. Such execution would require some implementation of complex functions or predicates. Since the OS may be a smooth abstraction of a family of solvers, it is in principle possible to develop some simulation producing supersets of possible traces. This can be useful to analyse some properties of the traces to improve their design. However it will become worth and more interesting when a more refined full trace will be ready.

\vspace{1mm}
Another point which remains unsolved is the way to relate the observational semantics of the tracer and the design of the CHROME-REF PIM. Both are forms of partial formal specification. The OS because it is an abstraction of the semantics of the observed process - hence a partial specification-, the later because it is a partially formal specification. Even if it is clear that the description of the tracer in SFC is a clear requirement which serves as guideline to design this PIM, there is no way to guarantee a formal correspondence. 

\begin{figure} 
\centering
\scalebox{0.9}{\includegraphics{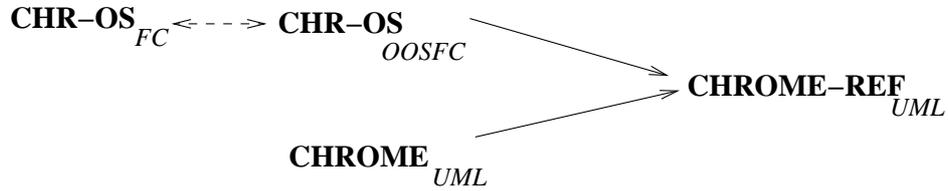}}
\caption[An Intermediate Step towards CHROME-REF] {An Intermediate Step towards CHROME-REF}
\label{fig:refined_approach_trace}
\end{figure}

One proposed approach is to map the logical model of SFC used here into an Object-Oriented (OO) model (OOSFC), and to limit the ``implementation step'' to a merging of PIMs. This way to proceed is illustrated on the Figure~\ref{fig:refined_approach_trace}.  This approach aims at reducing
the complexity of mapping between the two different descriptions, by introducing a intermediary step denoted CHR-OS$_{OOSFC}$.

We have started to specify the OS in OOSFC \cite{Rafael09a}, but this question is still open whether this step is really helping, or whether the construction of the tracer part of the CHROME-REF PIM in UML is better achieved just using the SFC specification of the OS. It could be also interesting to compare several approaches of extending existing codes, like pluging aspects in the CHROME java code. In any cases the question of the relationships with the specification is still worth posing.

\begin{figure} 
\centering
\scalebox{0.9}{\includegraphics{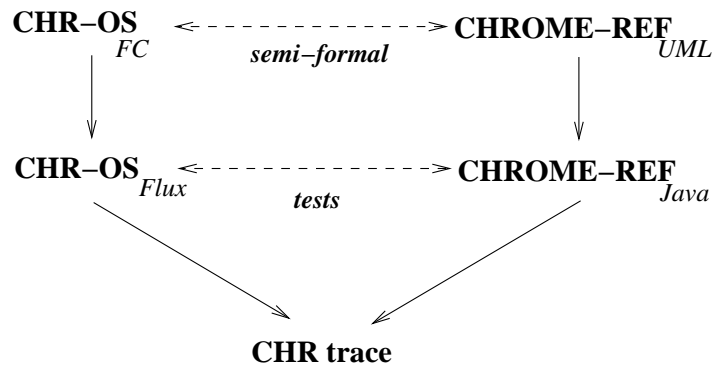}}
\caption[Validating different operational semantics] {Validating different operational semantics}
\label{fig:validation_trace}
\end{figure}

\vspace{1mm}
Finally a third unsolved point concerns the validation of the implementation. Considering the design and implementation used method, there is no way to make formal proof of adequation between the specification and the implementation, but this point need more study. At this stage, we are limited to perform tests. The Figure~\ref{fig:validation_trace} illustrates this point.

\begin{itemize}
  \item \textbf{Semi-formal proof}: It is to show that CHR-OS$_{FC}$ is equivalent to the CHROME-REF$_{UML}$. But, as UML is a semi-formal language
  \cite{latella1towards}, so that, UML is not stricly formal in sense of a
  purely syntactic derivation using a very precise and circumscribed formal set
  of rules of inference, no formal proof can be performed.
  \item \textbf{Test-based validation}: Let CHR-OS$_{FC}$ with its
  equivalent implementation CHR-OS$_{Flux}$ in Flux,  and CHROME-REF$_{UML}$ with its equivalent
  implementation in Java CHROME-REF$_{Java}$, the method consists of
  comparing the produced traces to check whether they are equivalent.
\end{itemize} 

\vspace{5mm}
There is still a long way to get a full generic trace for CHR. Indeed, as quoted in the Section~\ref{sec:compoCHRsem}, such goal implies the existence of generic traces for several CHR extensions. Since we already have some (for finite domain solvers, or for Prolog under-layer for example), we are far to cover all existing extensions of CHR quoted at the beginning of this paper. But the composition of traces and the method of implementing tracer presented here give a possible road towards a full generic trace for CHR$^\vee$ and many of its extensions.


\newpage

\nocite{*}
\bibliographystyle{acm}  
\bibliography{report}  

\begin{appendix}
\section{Appendix - Observational Semantics of CHR}
\label{app:obssemCHRor} 

The following is the description of the observational semantics of CHR based on its theoretical operational semantics $\omega_t$ and the simple fluent calculus with modified axioms of the Section~\ref{sec:obssemSfc}.

\vspace{3mm}
\begin{enumerate}[(a)]
  \item Domain Sorts
  \begin{enumerate}[-]
	  \item $NATURAL$, natural numbers;
	  \item $RULE$, the sort of CHR rules and $RULE\_ID$ the sort of the rule
	  identifiers;
	  \item $CONSTRAINT$, the sort of constraints, with the following subsorts:
	  $BIC$ (the built-in constraints), with the subsort $EQ$ (constraints in the
	  form $x = y$), and $UDC$ (the user-defined constraints), with the following
	  subsort: $IDENTIFIED$ (constraints in the form $c \# i$). In short: 
          
          $EQ < BIC  < CONSTRAINT$ and 

	  $IDENTIFIED < UDC < CONSTRAINT$;
	  \item $PROPHISTORY = Seq(NATURAL) \times RULE$, the elements of the
	  Propagation History, tuples of a sequence of natural numbers and a rule.
	  For each defined sort $X$, three new sorts: $Seq(X)$, $Set(X)$ and
	  $Bag(X)$ containing the sequences, the sets and the bags of elements of $X$.
	  We use [] for the empty sequence and \{\} for the empty set and the empty bag.  
	  \item $CHR ACTION < ACTION$, the subsort of $ACTION$ containing only the
	  actions in the CHR semantics.    
  \end{enumerate}

  \item Predicates
  \begin{enumerate}[-]
	  \item $Query : Bag(CONSTRAINT)$, $Query(q)$ holds iff $q$ is the initial
	  query;
	  \item $Consistent : STATE$, holds iff the $BICS$ of the state is consistent
	  (i.e., if it does not entail false);
	  \item $Match(h_k, h_R, u_1, u_2, e, z)$ holds iff (i) $u_1$ and
	  $u_2$ are in the $UDCS$ of $z$ and (ii) the set of matching equations e is
	  such that $chr(u_1) = e(h_k)$ and $chr(u_2) = e(h_R)$;
	  \item $Entails : Set(BIC) \times Set(EQ) \times Bag(BIC)$, $Entails(b, e, g)$
	  holds if $CT \models b \rightarrow \exists (e \wedge g)$.
  \end{enumerate}  

  \item Functions
  \begin{enumerate}[-]
	  \item $\# : UDC \times NATURAL \mapsto IDENTIFIED$, defines the syntactic
	  sugar for defining identified constraints in the form $c\#i$;   
	  \item $makeRule : $

$	  RULE ID \times Bag(UDC) \times Bag(UDC) \times
	  Bag(BIC) \times Bag(UDC) \mapsto RULE$, makes a rule from its components. We
	  define the syntactic sugar for rules as $r_{id} @ h_k \backslash h_R
	  \leftrightarrow g|b = makeRule(r_{id}, h_k, h_R, g, b)$;
	  \item $Bics : STATE \mapsto Set(BIC)$, where $Bics(z) =
	  \left\{c|Holds(InBics(c), z)\right\}$;
	  \item $id : Set(UDC) \mapsto Set(NATURAL)$, where $id(H) = {i|c\#i \in H}$
	  \item The usual set, sequence and bag operations: $\in$ for pertinence, $\cup$
	  for set union, $\uplus$ for bag union, $++$ for sequence concatenation, $|$
	  for sequence head and tail (Ex: $ [ head|tail] $) and $ \backslash $ for set 
	  subtraction.
  \end{enumerate}    

  \item Fluents
  \begin{enumerate}[-]
	  \item $Query : Bag(UDC) \mapsto FLUENT$, $Query(q)$ holds iff $q$ is the
	  initial toplevel goal;
	  \item $Goal : Bag(UDC) \mapsto FLUENT$, $Goal(q)$ holds iff $q$ is the
	  current goal;
	  \item $Udcs : Bag(IDENTIFIED) \mapsto FLUENT$, $Udcs(u)$ holds iff $u$ is the
	  current UDCS;
	  \item $InBics : BIC \mapsto FLUENT$, $InBics(c)$ holds iff $c$ is in the
	  current BICS;
	  \item $InPropHistory : PROPHISTORY \mapsto FLUENT$, $InPropHistory(p)$ holds
	  iff $p$ is in the current Propagation History;
	  \item $NextId : NATURAL \mapsto FLUENT$, $NextId(n)$ holds iff $n$ is the
	  next natural number to be used to identify a identified constraint.    
  \end{enumerate} 

  \item Actions 
  \begin{enumerate}[-]
          \item $Init : \mapsto CHR\_ACTION$, $Do(Init, [goal(q)|a], s)$ executes the
	  toplevel initial transition (starting the resolution) with some query $q$ in the current state ($a$ stands for other attributes list in the associated trace event);
	  \item $Solve : \mapsto CHR\_ACTION$, $Do(Solve,[bic(c)|a], s)$ executes the
	  $Solve$ transition with the built-in constraint $c$;
	  \item $Introduce : \mapsto CHR\_ACTION$, $Do(Introduce, [udc(c)|a], s)$ executes
	  the Introduce transition with the user-defined constraint $c$;
	  \item $Apply : \mapsto CHR\_ACTION$,
	  
           $Do(Apply, [rule(r)|t], s)$ executes the Apply transition with rule $r$ matching the
	  constraints in the $UDCS$ with the kept and removed heads;
	  \item $Fail : \mapsto CHR\_ACTION$, $Do(Init, [goal(q)|a], s)$ executes the
	  toplevel initial transition (starting the resolution) with some query $q$ in the current state ($a$ stands for other attributes list in the associated trace event);
  \end{enumerate}

\item Attributes
  \begin{enumerate}[-]
\item $goal : CONSTRAINTS \mapsto ATTRIBUTE$, is the set of constraints in the current Goal;
	  \item $udc : CONSTRAINTS \mapsto ATTRIBUTE$, is the set of constraints in the current User Defined Constraints Store;
	  \item $bic : CONSTRAINTS \mapsto ATTRIBUTE$, is the set of constraints in the current Built-In Constraints Store ;
	  \item $hind : \mapsto INTEGER$, is the new propagation history index (incremented by $Introduce$);
	  \item $rule: RULE \mapsto ATTRIBUTE$, is the rule applied to reach this state.
  \end{enumerate}

\newpage
  \item {\bf Axioms of the Observational Semantics}

\vspace{4mm}
\texttt{Init}

  \begin{enumerate}[-]
	  \item $Poss(Init, [q], z) \equiv Holds(Query(q), z)$ 

	  \item $Poss(Init, [q], State(s)) \supset State(Do(Init, $[initState$, goal(q), hind(1)$]$, s)) = $

$\ \ \ \ \ \ State(s) \circ Udcs(\left\{\right\}) \circ Goal(q) \circ NextId(1) - $

$\ \ \ \ \ \ Query(q)$

	  The Initial State Axiom states that in the initial state,
	  the goal contains the constraints in the query, the user defined constraint
	  store is empty and the next ID for identified constraints is 1;

\vspace{4mm}
\texttt{Solve}

	  \item $Poss(Solve, [c], z) \equiv (\exists q)(Holds(Goal(q \uplus \left\{c\right\}), z)$
	 
	 The Solve Precondition Axiom states that the only precondition for the
	 $Solve$ action on the built-in constraint $c$ is that this constraint should
	 be in the goal.

	 \item $Poss(Solve, [c], s) \supset  $

$\ \ \ \ State(Do(Solve,$\  [solve$, bic(c), goal(q)$]$, s)) = $

$\ \ \ \ \ \ State(s) \circ Goal(q) \circ InBics(c) -  Goal(q \uplus \left\{c\right\})$ 
	
	The Solve State Update Axiom states that the result of the $Solve$ action over
	the constraint $c$ is that this constraint is removed from goal and added to
	$InBics$ list in current state;

\vspace{4mm}
\texttt{Introduce}

	\item $Poss(Introduce, [c], z) \equiv (\exists q)(Holds(Goal(q), z) \wedge c \in q)$

	\item $Poss(Introduce, [c], State(s)) \wedge Holds(Udcs(u), State(s)) \wedge $

$\ \ \ \ \ \ Holds(NextId(n), State(s)) \supset $

$\ \ \ \ State(Do(Introduce, $\ [introduce$, udc(c), goal(q), hind(n+1)$]$, s)) = $

$\ \ \ \ \ \ State(s) \circ Goal(q) \circ Udcs(u \uplus {c\#n}) \circ NextId(n + 1) - $

$\ \ \ \ \ \ Goal(q \uplus \left\{c\right\}) - Udcs(u) - NextId(n)$

\vspace{4mm}
\texttt{Apply}
 
    \item $Poss(Apply, [r, h_k, h_R, g, u_1, u_2], z)  \equiv $

$\ \ \ \ (\exists e)(\exists b)(Match(h_k, h_R, u_1, u_2, e, z) \wedge$
    
$\ \ \ \ \lnot Holds(InPropHistory(id(u_1), id(u_2), r), z) \wedge Bics(b, z) \wedge $

$\ \ \ \  Entails(b, e, g))$

    \item $Poss(Apply, [r, h_k, h_R, g, u_1, u_2], State(s)) \wedge $

$\ \ \ \ Holds(Udcs(u_1 \uplus u_2 \uplus u), State(s))
\wedge Holds(Goal(q), State(s)) \wedge $

$\ \ \ \ Match(h_k, h_R, u_1, u_2, e, z) \supset $

$\ \ \ \ State(Do(Apply, $

\ \ \ \ [apply, $rule(r @ h_k \backslash h_R \leftrightarrow g|d, u_1, u_2), goal(d \uplus q), udc(u1 \uplus u), bic(g)$]$, s)) =$

$\ \ \ \ State(s) \circ Goal(d \uplus q) \circ Udcs(u1 \uplus u) \circ InBics(e) \circ InBics(g) \circ $

$\ \ \ \ \ \  InPropHistory(id(u_1), id(u_2), r) $

$\ \ \ \ \ \  - Goal(q) - Udcs(u1 \uplus u2 \uplus u)$

\vspace{4mm}
\texttt{Fail}

	  \item $Poss(Fail, [q], z) \equiv Holds(goal(q),z) \wedge \not \exists Poss(Apply, [r, h_k, h_R, g, u_1, u_2], z)$
	 
	 \item $Poss(Fail, [q], s) \supset $

$\ \ \ \ State(Do(Fail, $\ [fail$, goal(q) $]$, s)) = State(s)$

  \end{enumerate}   
\end{enumerate}

\section{Appendix - XML schema for Generic CHR Trace}
\label{app:chrtraceschema} 

\lstset{
	backgroundcolor==\color{darkgray},
	tabsize=4,
	rulecolor=,
	language=xml,
        basicstyle=\footnotesize,  
        upquote=true,
        aboveskip={1.5\baselineskip},
        columns=fixed,
        showstringspaces=false,
        extendedchars=true,
        breaklines=true,
        prebreak = \raisebox{0ex}[0ex][0ex]{\ensuremath{\hookleftarrow}},
        showtabs=false,
	numberstyle=\tiny,
	numbersep=5pt,
    showspaces=false,	
    showstringspaces=false,
    identifierstyle=\ttfamily,
    keywordstyle=\color{black}\bfseries\emph,
    commentstyle=\color[rgb]{0.133,0.545,0.133},
    stringstyle=\color[rgb]{0.627,0.126,0.941},
}
\begin{lstlisting}
<?xml version="1.0" encoding="UTF-8"?>
 <xs:schema xmlns:xs="http://www.w3.org/2001/XMLSchema"
   targetNamespace="http://orcas.org.br/chrv" xmlns="http://orcas.org.br/chrv"
   elementFormDefault="qualified">
  <xs:element name="chrv"> 
   <xs:complexType>
    <xs:sequence>
     <xs:element name="event" minOccurs="0" maxOccurs="unbounded">
      <xs:complexType>
       <xs:choice>
        <xs:element name="initialState" minOccurs="1" maxOccurs="1">
         <xs:complexType>
          <xs:sequence>
           <xs:element name="goal" type="xs:string" />
           <xs:element name="hind" type="xs:integer" />
         </xs:sequence>
        </xs:complexType>
       </xs:element>
        <xs:element name="introduce" minOccurs="1" maxOccurs="1">
         <xs:complexType>
          <xs:sequence>
           <xs:element name="udc" type="xs:string" />
           <xs:element name="goal" type="xs:string" />
           <xs:element name="hind" type="xs:integer" />
         </xs:sequence>
        </xs:complexType>
       </xs:element>
        <xs:element name="solve" minOccurs="1" maxOccurs="1">
         <xs:complexType>
          <xs:sequence>
           <xs:element name="bic" type="xs:string" />
           <xs:element name="goal" type="xs:string" />
         </xs:sequence>
        </xs:complexType>
       </xs:element>
        <xs:element name="apply" minOccurs="1" maxOccurs="1">
         <xs:complexType>
          <xs:sequence>
           <xs:element name="rule" type="xs:string" />
           <xs:element name="goal" type="xs:string" />
           <xs:element name="udc" type="xs:string" />
           <xs:element name="bic" type="xs:string" />
          </xs:sequence>
         </xs:complexType>
        </xs:element>
         <xs:element name="fail" minOccurs="1" maxOccurs="1">
          <xs:complexType>
           <xs:sequence>
            <xs:element name="rule" type="xs:string" />
           </xs:sequence>
          </xs:complexType>
         </xs:element>
        </xs:choice>
         <xs:attribute name="chrono" type="xs:string" use="required" />
       </xs:complexType>
      </xs:element>
     </xs:sequence>
    </xs:complexType>
     <xs:unique name="chronoKey" />
     <xs:selector xpath="event" />
     <xs:field xpath="@chrono" />
   </xs:unique>
  </xs:element>
 </xs:schema>
\end{lstlisting}



\section{Appendix - LEQ Example Execution Trace}
\label{app:leqexample}

Here is the trace of execution of the LEQ the Example 3.1 executed in CHROME-REF. 

\begin{lstlisting}
<?xml version="1.0" encoding="UTF-8"?>
<chrv
	xmlns="http://orcas.org.br/chrv"
	xmlns:xsi="http://www.w3.org/2001/XMLSchema-instance" 
	xsi:schemaLocation=
		"http://orcas.org.br/chrv chrv2.xsd">	 
	<event chrono="1">
		<initialState>
			<goal> leq(A,B), leq(B,C), leq(C,A) </goal>
			<hind> 1 </hind>
		</initialState>
	</event>
	<event chrono="2">
		<introduce>
			<udc> leq(A,B) </udc>
			<goal> leq(B,C), leq(C,A) </goal>
			<hind> 2 </hind>
		</introduce>
	</event>
	<event chrono="3">
		<introduce>
			<udc> leq(A,B), leq(B,C) </udc>
			<goal> leq(C,A)) </goal>
			<hind> 3 </hind>
		</introduce>
	</event>
	<event chrono="4">
		<apply>
			<rule> r4@ leq(A,B), leq(B,C) ==> leq(A,C) </rule>
			<goal> leq(C,A), leq(A,C) </goal>
		</apply>
	</event>
	<event chrono="5">
		<introduce>
			<udc> leq(A,B), leq(B,C), leq(A,C) </udc>
			<goal>leq(C,A)</goal>
			<hind> 4 </hind>
		</introduce>
	</event>
	<event chrono="6">
		<introduce>
			<udc> leq(A,B), leq(B,C), leq(A,C), leq(C,A) </udc>
			<goal> </goal>
			<hind> 5 </hind>
		</introduce>
	</event>
	<event chrono="7">
		<apply>
			<rule> r2@ leq(A,C), leq(C,A) ==> A=C </rule>
			<goal> </goal>
			<udc> leq(C,B), leq(B,C) </udc>
			<bic> A=C </bic>
		</apply>
	</event>
	<event chrono="8">
		<apply>
			<rule> r2@ leq(C,B), leq(B,C) ==> C=B </rule>
			<goal> </goal>
			<udc> </udc>
			<bic> A=C, C=B </bic>
		</apply>
	</event>
</chrv>
\end{lstlisting}

No more LEQ program rule may apply.

\vspace{5mm}
Using a representation where attributes have the functional form used in the Observational Semantics, it corresponds to the trace (attributes with empty argument are omitted):

\vspace{4mm}
See Appendix~\ref{app:obssemCHRor} for the meaning of the attributes.

\begin{verbatim}
 1    initialState goal((leq(A,B), leq(B,C), leq(C,A))))
                   hind(1)
 
 2    introduce    udc((leq(A,B)))
                   goal((leq(B,C), leq(C,A)))
                   hind(2)

 3    introduce    udc((leq(A,B), leq(B,C)))
                   goal((leq(C,A)))
                   hind(3)        

 4    apply        rule((r4@ leq(A,B), leq(B,C) ==> leq(A,C)))
                   goal((leq(C,A), leq(A,C)))

 5    introduce    udc((leq(A,B), leq(B,C), leq(A,C))) 
                   goal((leq(C,B)))
                   hind(4)    

 6    introduce    udc((leq(A,B), leq(B,C), leq(A,C), leq(C,A)))
                   hind(5)

 7    apply        rule((r2@ leq(A,C), leq(C,A) ==> A=C)), 
                   udc(leq(C,B), leq(B,C))    
                   bic((A=C))

 8    apply        rule((r2@ leq(C,B), leq(B,C) ==> C=B)), 
                   bic((A=C, C=B ))
\end{verbatim}   
\section{Appendix - OS of a Robots Application in SFC}
\label{app:robots} 


This world consists of agents (the robots) moving in a space structured by rooms connected by  doors, able to carry objects they find in the rooms. A requisition is an order to seek for objects and carry them from some place to an other one. The scene description at some moment is the current state and consists of a set of facts. A ``situation'' corresponds to a succession of trace events. The current state corresponding to a given situation is obtained here by the reconstruction function (interpretation semantics).

In fluent calculus, facts are named ``fluents'' and
requisitions (or requests) are similar to Prolog goals. 
The way the requisitions are computed is not described by the observational semantics. The requests are thus treated as influence factors.

\vspace{2mm}
We describe a simplified version of the example of \cite{Thielscher98} with 3 rooms.

\begin{figure} [h]
\begin{center}
\includegraphics[width=0.25\linewidth]{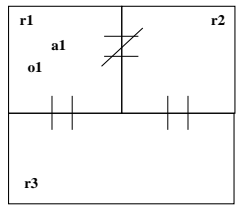}
\end{center}
\caption{A simple robot world}
\label{figcontext}
\end{figure}

The simplified robot's world is depicted on Figure~\ref{figcontext}.

Initially there is one object and one robot both located in the same room, and the door \verb.d12. is locked. The robot has its key.

We present an implementation of the OS in Flux.
A current state is described by a set of atoms.

\vspace{5mm}
{\bf \underline{ACTIONS TYPES} }

\vspace{1mm}
{\bf pickup} pick an object (if any)

{\bf drop} drop the carried object (if any)

{\bf gotodoor} go to the quoted door (if any)

{\bf enteroom}  enter the quoted room (if the door is open)

{\bf open} open the door (if it is closed)

\vspace{5mm}
{\bf \underline{TRACE EVENTS} [attributes]}

\vspace{1mm} 
Attribute \verb.a. stands for ``agent''

Attribute \verb.o. stands for ``object''

Attribute \verb.r. stands for ``room''

Attribute \verb.d. stands for ``door''

\vspace{1mm}
\begin{verbatim}
pickup a o r
drop   a o r
walk   a d
walk   a r
open   a d
\end{verbatim}

\begin{itemize}
\item Domains

\vspace{3mm}

\begin{tabular}{|l|l|}
  \hline
  \multicolumn{2}{|c|}{Domain Sorts} \\
  \hline
  {\scriptsize AGENT} & $A1$ \\
  {\scriptsize ROOM} & $R1, R2, R3$ \\
  {\scriptsize DOOR} & $D12, D13$ \\
  {\scriptsize OBJECT} & $O1, O2, O3$ \\
  \hline
\end{tabular}   

\vspace{3mm}
\item Parameters

\vspace{3mm}
\scalebox{0.8}{
\begin{tabular}{|l|l|l|}
  \hline
  \multicolumn{3}{|c|}{Parameters versus Fluents} \\
  \hline
parameter & type & meaning \\ \hline
\textit{AgentInRoom} & {\scriptsize AGENT $\times$ ROOM $\mapsto$ FLUENT} & the agent is in room $r$ \\
\textit{AtDoor} & {\scriptsize AGENT $\times$ DOOR $\mapsto$ FLUENT} & the agent is at door $d$ \\
\textit{Closed} & {\scriptsize DOOR $\mapsto$ FLUENT} & door $d$ is closed \\
\textit{Carries} & {\scriptsize AGENT $\times$ OBJECT $\mapsto$ FLUENT} & agent carries object $o$ \\
\textit{HasKeyCode} & {\scriptsize AGENT $\times$ DOOR $\mapsto$ FLUENT} & agent has the key code for door $d$ \\
\textit{ObjectInRoom} & {\scriptsize OBJECT $\times$ ROOM $\mapsto$ FLUENT} & the object is in room $r$ \\
\textit{Request} & {\scriptsize ROOM $\times$ OBJECT $\times$ ROOM $\mapsto$ FLUENT} & there is a request to deliver \\
& & object $o$ from room $r_1$ to room $r2$ \\
\hline
\end{tabular} 
}

\vspace{3mm}
Each parameter may be represented by several fluents.
({\em Request} is treated as external)

\vspace{4mm}
The initial state is formalized by this term below.

\vspace{3mm}

$Holds(AgentInRoom(A1, R2), S_0) \land Holds(ObjectInRoom(O1, R3), S_0) \land \\
Holds(ObjectInRoom(O2, R1), S_0) \land Holds(ObjectInRoom(O3, R2), S_0) \land \\
Holds(Closed(D12), S_0) \land Holds(HasKeyCode(A1, D13), S_0) \land \\
Holds(Request(R3, O1, R2), S_0) \land Holds(Request(R1, O2, R3), S_0) \land \\
Holds(Request(R2, O3, R1), S_0) \land (\forall x)\lnot Holds(Carries(A1, x), S_0) \\$ 

\vspace{3mm}
\item Auxiliary Predicates

\vspace{3mm}
\begin{tabular}{|l|l|l|}
  \hline
  \multicolumn{3}{|c|}{Auxiliary Predicates} \\
  \hline
predicate & type & meaning \\ \hline
\textit{Connects} & {\scriptsize ROOM $\times$ DOOR $\times$
ROOM} & door $d$ connects rooms $r_1$ and $r_2$  \\
\hline
\end{tabular} 

\item Actions and Actual state Attributes

\vspace{2mm}
\begin{tabular}{|l|l|l|}
  \hline
  \multicolumn{3}{|c|}{Actions} \\
  \hline
action & attributes & action meaning \\ \hline
\textit{Pickup} & {\scriptsize pickup $\times$ AGENT $\times$ OBJECT $\times$ ROOM } & pick up object $o$ \\ 
\textit{Drop} & {\scriptsize drop $\times$ AGENT $\times$ DOOR $\times$ ROOM } & drop object $o$ \\ 
\textit{GoToDoor} & {\scriptsize walk $\times$ AGENT $\times$ DOOR } & go to door $d$ \\ 
\textit{EnterRoom} & {\scriptsize walk $\times$ AGENT $\times$ ROOM} & enter room $r$ \\ 
\textit{Open} & {\scriptsize open $\times$ DOOR} & open door $d$ \\
\hline
\end{tabular}

\end{itemize}

\vspace{3mm}
The parameters of the actions in a condition are just used for communication of particular values and thus avoid rewriting of ``Holds'' conditions in the following axiom.

\vspace{5mm}
\textbf{Observational Semantics}

\vspace{5mm} 
\texttt{Pickup}

$Poss(Pickup, [a,o,r], s) \equiv $

$\ \ \ \ \ Holds(AgentInRoom(a, r), s) \land Holds(ObjectInRoom(o, r), s) \land $

$\ \ \ \ \ \lnot Holds(Carries(a, o))) \\$

$Poss(Pickup, [a,o,r], s) \supset $

$\ \ \ \ \ State(Do(Pickup, [pickup, a, o, r], s)) = State(s) \circ Carries(a, o)\\$

\vspace{2mm}
\texttt{Drop} 

$Poss(Drop, [a,o,r], s) \equiv $

$\ \ \ \ \ Holds(Carries(a, o)) \land Holds(AgentInRoom(a, r), s) \\$

$Poss(Drop, [a,o,r], s) \supset $

$\ \ \ \ \ State(Do(Drop, [drop, a, o, r], s)) = State(s) - Carries(a, o)\\$

\vspace{2mm}
\texttt{GoToDoor}

$Poss(GoToDoor, [a, d, r], s) \equiv Holds(AgentInRoom(a, r), s) \land $

$\ \ \ \ \ (\exists r') Connects(r, d, r') \land \lnot(\exists d') Holds(AtDoor(a, d'), s)\\$

$Poss(GoToDoor, [a, d, r], s) \supset $

$\ \ \ \ \ State(Do(GoToDoor, [walk, a, d], s)) = State(s) \circ AtDoor(a, d)\\$

\vspace{2mm}
\texttt{EnterRoom}

$Poss(EnterRoom, [a, r, d, r'], s) \equiv Holds(AgentInRoom(a, r)) \land $

$\ \ \ \ \ Holds(AtDoor(a, d), s) \land Connects(r, d, r') \land \lnot Holds(Closed(d), s)) \\$

$Poss(EnterRoom, [a, r, d, r'], s) \supset $

$\ \ \ \ \ State(Do(EnterRoom, [walk, a, r'], s)) = State(s) \circ AgentInRoom(a, r') -$

$\ \ \ \ \ AgentInRoom(a, r)\\$

\vspace{2mm}
\texttt{Open}

$Poss(Open, [a, d], s) \equiv $

$\ \ \ \ \ Holds(AtDoor(a, d), s) \land Holds(HasKeyCode(a, d), s) \land $

$\ \ \ \ \ Holds(Closed(d),s)\\$

$Poss(Open, [a, d], s) \supset $

$\ \ \ \ \ State(Do(Open, [open, a, d], s)) = State(s) - Closed(d)\\$


\newpage
\textbf{Example of trace:}

\begin{verbatim}
 1   pickup   a1   o1   r1
 2   walk     a1   d12 
 3   open     a1   d12
 4   walk     a1   r2 
 5   walk     a1   d12 
 6   walk     a1   r1
 7   drop     a1   o1   r1 
 8   pickup   a1   o1   r1 
 9   drop     a1   o1   r1 
10   walk     a1   d13
\end{verbatim}

   
\end{appendix}         

\newpage
\tableofcontents

\end{document}